\documentclass[useAMS,usenatbib]{mn2e}
\usepackage{amsmath,amsfonts,amssymb}
\usepackage{}
\usepackage{rotating}  
\usepackage{theorem}   
\usepackage{lscape}
\usepackage{rotfloat}
\usepackage{longtable}
\usepackage{pdflscape}
\usepackage{float}
\usepackage{longtable}
\usepackage{multirow}
\usepackage{bm}  
\usepackage{booktabs}  
\usepackage{pdfsync}
\usepackage{threeparttable}   
\usepackage{listings}
\usepackage{mathtools} 
\usepackage{xcolor}
\usepackage{natbib}

\def\href#1#2{#2}
\def\urlinner#1{#1\endgroup}
\def\url{\begingroup\def\do##1{\catcode`##1 12 }%
  \do\\\do\$\do\&\do\#\do\^\do\_\do\%\do\~ \ttfamily \urlinner}

\newcommand{\msun}{M$_\odot$}
\newcommand{\mjup}{M$_{\rm J}$}
\newcommand{\rsun}{R$_\odot$}

\newcommand{\mytilde}{\raise.17ex\hbox{$\scriptstyle\mathtt{\sim}$}}
\def\lesssim{\mathrel{\hbox{\rlap{\hbox{\lower3pt\hbox{$\sim$}}}\hbox{\raise2pt\hbox{$<$}}}}}
\def\gtrsim{\mathrel{\hbox{\rlap{\hbox{\lower3pt\hbox{$\sim$}}}\hbox{\raise2pt\hbox{$>$}}}}}
\def\lesseq{\mathrel{\hbox{\rlap{\hbox{\lower3pt\hbox{$-$}}}\hbox{\raise2pt\hbox{$<$}}}}}
\def\gtreq{\mathrel{\hbox{\rlap{\hbox{\lower3pt\hbox{$-$}}}\hbox{\raise2pt\hbox{$>$}}}}}

\title[Tides and the PN Binary Population]{The Effect of Tides on the Population of PN from Interacting Binaries}
\author[N. Madappatt, O. De Marco \& E. Villaver]{Niyas Madappatt$^{1,2,3}$\thanks{E-mail:niyas.alikutty@mq.edu.au}, Orsola De Marco$^{1,2}$ and  Eva Villaver $^{4}$
\\$^{1}$Department of Physics and Astronomy , 
           Macquarie University,  
           North Ryde,  
           Sydney,  
          Australia\\
$^{2}$Astronomy, Astrophysics and Astrophotonics Research Centre, 
           Macquarie University,  
           North Ryde,  
           Sydney,  
          Australia\\
$^{3}$School of Computing, Engineering and Mathematics, 
           Western Sydney University,  
           Penrith,   
          Australia\\
$^4$Universidad Aut\'{o}noma de Madrid,
Campus de Cantoblanco
Fco.Tom\'{a}s y Valiente, 7
28049 Madrid, 
Spain}      
          
\begin{document}

\maketitle

\label{firstpage}

\begin{abstract}
We have used the tidal equations of Zahn to determine the maximum orbital distance at which companions are brought into Roche lobe contact with their giant primary, when the primary expands during the giant phases. This is a key step when determining the rates of interaction between giants and their companions. Our stellar structure calculations are presented as maximum radii reached during the red and asymptotic giant branch (RGB and AGB, respectively) stages of evolution for masses between 0.8 and 4.0~\msun\  (Z=0.001 -- 0.04) and compared with other models to gauge the uncertainty on radii deriving from details of these calculations. We find overall tidal capture distances that are typically 1-4 times the maximum radial extent of the giant star, where companions are in the mass range from 1~M$_{\rm J}$ to a mass slightly smaller than the mass of the primary. We find that only companions at initial orbital separations between $\sim$320 and $\sim$630~\rsun\ will be typically captured into a Roche lobe-filling interaction or a common envelope on the AGB. Comparing these limits with the period distribution for binaries that will make PN, we deduce that in the standard scenario where all $\sim$1-8~\msun\ stars make a PN, at most 2.5 per cent of all PN should have a post-common envelope central star binary, at odds with the observational lower limit of 15-20 per cent. The observed over-abundance of post-interaction central stars of PN cannot be easily explained considering the uncertainties. We examine a range of explanations for this discrepancy.
\end{abstract}


\section{Introduction}
\label{sec:introduction}

During the late evolutionary stages of a star, a planetary nebula (PN) is formed. PN with shapes diverging  from spherical account for $\sim$80 per cent of the entire population (\citealt[][]{Parker2006}). So far, single stellar evolution theory fails to explain PN shapes, except for spherical and mildly elliptical ones \citep{Soker2006,Nordhaus2007,GarciaSegura2014}. Interactions with companions can trivially explain non-spherical shapes \citep{Soker1997,Mitchell2007,Miszalski2009b,DeMarco2011}.  A series of observational efforts have been trying to determine the impact of binarity on PN formation and shaping \citep{Bond2000,Miszalski2009,Jones2014,DeMarco2009,DeMarco2013,Douchin2015}.

Binary interactions, including interaction with a planetary system, can alter the stellar mass-loss rate and/or the geometry of the outflow when stars become giants, either during the red or asymptotic giant branch (RGB and AGB) phases. In wider binary interactions, a companion can focus or accrete from the wind of a giant star. If the orbital separation is shorter, the companion comes into contact with the giant when the latter fills its Roche lobe \citep{Soker1997}, mass transfer takes place and this can result, if the mass transfer is unstable, in a common envelope (CE) interaction \citep{Paczynski1976,Ivanova2013}. In all cases, tidal forces between the companion and the expanding giant will extend the initial orbital separation for which an interaction will eventually take place. 

The fraction of PN with post-CE central binaries must be influenced by the action of tidal forces that extend the ability of the RGB or AGB star to draw in an orbiting companion and ``capture"\footnote{The term ``capture" is not formally correct, because the companions are already in bound orbits. However, it is a convenient term that gives the idea that the giant star brings in the companion, ``capturing" it into an interaction.} it into an interaction. Once captured, these companions force the primary to fill its own Roche lobe. If the primary is an AGB giant and unstable mass transfer takes place, the two stars will in-spiral \citep{Paczynski1976} and the AGB star's envelope may be ejected as a result.

At least 15-20 per cent of all PN with stars bright enough to be monitored photometrically with small aperture telescopes have post-CE, close binaries in their centres \citep{Bond2000,Miszalski2009}. This fraction is a lower limit, because only the closest binaries can be detected by the adopted survey method, but the actual, unbiased fraction is likely not much larger than this \citep{DeMarco2008,DeMarco2015}. In the standard PN evolutionary scenario where all (single and binary) $\sim$1-8~\msun\ stars make a PN, this fraction must be consistent with the progenitor population binary fraction and its period distribution \citep[both of which are well characterised;][]{Duquennoy1991,Raghavan2010}. 

A back of the envelope calculation already shows that there are too few close binaries among main sequence progenitors of PN to result in the observed incidence of post-CE binaries in the PN population. This already implies that some stars, for example single stars, could make a sub-luminous PN, which are not readily observed \citep{Soker2005}. This would inflate the fraction of PN with binary central stars.  Observations may already be hinting at the fact that there is a hidden population of very faint spherical PN \citep{Jacoby2010}.

To go beyond a back of the envelope calculation and carry out an actual prediction we need to know, critically, how far out RGB and AGB stars can capture companions. \citet{Soker1994c} and \citet{Soker1996} derived relatively large tidal maximum capture distances\footnote{Similarly to our use of the term ``capture" we will use ``maximum capture distance" to mean the largest main sequence orbital separation of a given binary that will result in a tidal capture, where capture is intended as the moment when the primary fills its Roche lobe. The maximum capture distance changes as a function of primary and companion mass and it refers to either the RGB or AGB: the  maximum capture distance for a capture on the RGB will be typically smaller than for the AGB.}, where the AGB star would capture companions that are 5-6 times farther than the AGB stellar radius. This would lead to a larger predicted fraction of post-CE PN. On the other hand, recent work on the orbital evolution of planets would suggest smaller maximum capture distances, closer to a couple of stellar radii \citep{Carlberg2009,Villaver2009,Mustill2012,Adams2013,Nordhaus2013,Penev2014}. These lower values would result in smaller predictions. 

The effects of tides on binary populations with stellar-mass companions was implemented by \citet{Hurley2002} in their Binary Stellar Evolution (BSE) population synthesis code. However, in that study stellar evolution, and most critically for tides, the stellar radius, was computed using fitting formulae, which did not account for the large radial excursions during the AGB thermal pulses. 

We here use the tidal formalism of \cite{Zahn1977,Zahn1989}, including the spin-orbit coupling equation in conjunction with the stellar radial evolution predicted by the code Modules for Experiments in Stellar Astrophysics (MESA; \citealt{Paxton2011,Paxton2013}) to predict the maximum orbital separations out to which AGB stars capture companions into a Roche lobe filling interaction.  

The structure of this paper is as follows. In Section~\ref{sec:Zahnnew}, we describe the physics and assumptions of our approach. In Section 3, we present the stellar evolution calculation used in this study. In Sections 4 we present the maximum capture distance for both the RGB and AGB phases and in Section~5 and we compare our results with previous work. Finally, we conclude and discuss in Section 6.

\maketitle

\section{The Tidal Equations } 
\label{sec:Zahnnew}

We use the following tidal equations for a star with a fully convective envelope orbited by a companion in a circular orbit  \citep{Zahn1977,Zahn1989} and describe below those aspects that have not been previously reported in the literature: 

\begin{equation}\label{eq:newaa1}
\begin{aligned}
\frac{1}{a}\frac{da}{dt}=& -\frac{4} {7}~\frac{f}{t_{f}}~\frac{M_{\rm{env}}}{M} ~q (1+q)\left(\frac{R}{a}\right)^8 \left(1- \frac {\Omega}{\omega}\right)\\
&- \frac{\dot{M}}{M+\,M_2}- \left(\frac{2\dot{M}_2}{M_2}+\frac{\dot{M}_2}{M+\,M_2}\right),
\end{aligned}
\end{equation}


\begin{equation}\label{eq:newOmega1}
\begin{aligned}
\frac{d}{dt}({I \, \Omega})=&\frac{2}{7}~\frac{f}{t_{f}}~q^2~\frac{M_{\rm{env}}}{M}~{M}~{R}^2~\left({{R}\over a}\right)^6\left( \omega-\Omega \right) 
 - \frac{2}{3}\dot{M} R^2 \Omega,
\end{aligned}
\end{equation}
 
\noindent where $M$, $M_{\rm env}$, $\dot{M}$, $R$, $\Omega$ and $I$ are the mass, envelope mass, mass-loss rate, radius, spin frequency and moment of inertia of the primary; $M_{\rm 2}$ and $\dot{M_2}$ are the mass of the companion and the mass accretion onto the companion, respectively; $a$, $\omega$ and $q$ are the orbital separation, orbital frequency and mass ratio ($M_{\rm 2}/M$) of the binary; $f$ and $t_{\rm f}$  are the tidal strength parameter and the turnover timescale (for the latter we have used equation 8 of \citet{Villaver2009}, shown to be accurate by \citet{Mustill2012}). The $f$ parameter is used instead of $\lambda_2$ ($f=21\lambda_2$) following \citet{Verbunt1995}; $f$ is proportional to $(\alpha/2)^{4/3}$ \citep{Zahn1989}, where $\alpha$ is the convective mixing length theory parameter. The value of $\alpha$ has been constrained to be between 1.6 and 2 \citep{Trampedach2011}. This therefore results in a value of $f$ close to unity. For our work, where we are considering orbital separations such that the strongest tidal interaction happens during the giant phases, a value of unity is reasonable during those phases \citep[][]{Goldreich1977,Rasio1996,Hurley2002}.

The primary star's mass-loss rate ($\dot{M}$), high during the giant phases, contributes to separating the binary and this is accounted for in the second term of the right hand side of Equation~\ref{eq:newaa1}. The value of $\dot{M}$ is therefore negative, making the second term in the right hand side of Equation~\ref{eq:newaa1} positive. Some of that mass is accreted by the companion ($\dot{M}_2$) and we account for this in the third term of the right hand side of Equation~\ref{eq:newaa1}, based on the formalism of \cite{Bondi1944} and equation 6 of \cite{Boffin1988}. { Such a treatment must be regarded as approximate, because the accretion onto the secondary is mediated by the formation of a disk whose viscous properties will lead to accretion onto the star. The generation of accretion disks and their properties are a subject of active study within the field of AGB binaries \citep[e.g.,][]{HuarteEspinosa2013}.}  
 
 In Equation~\ref{eq:newOmega1} we account for the spin-orbit interaction. The angular momentum lost from the system is accounted for in the second term of the right hand side by assuming that the star is losing mass uniformly from an outer shell, so the change in spin angular momentum due to mass loss is given by:
${\rm{\dot{J}_{ML}}}=\, \frac{2}{3} \dot{M} R^2 \Omega $, where the moment of inertia of a spherical shell is $\frac{2}{3}\,M\,R^2$. {We have checked that this approximation is similar to recalculating the star's moment of inertia using the time-dependent stellar density distribution. Using Equation~\ref{eq:newOmega1}, the angular momentum of the system is conserved to below one per cent.

In this work, as in other models of binaries subject to tides \citep[e.g., see][]{Izzard2009}, we do not include the companion-induced mass loss of \cite{Tout1988}. As pointed out by \cite{Hurley2002}, such enhanced mass-loss rates may not be realistic for a wide range of binary systems. { It is likely that some form of companion induced mass-loss is present as we discuss in Section~\ref{sec:mesa}, so this omission contributes to the list of uncertainties in a tidal calculation like this.}

We do not consider the tide induced by the primary on the secondary, because we consider exclusively compact companions such as main sequence or white dwarf stars. These are invariably more compact than the primary during the phases of strong tidal interaction when the primary is a giant. It is highly unlikely that both stars in the binary are giants at the same time because if so their initial masses would have be the same to within a few percent, something that is unlikely in the PN progenitor population \citep{Raghavan2010}. 

We evolve the orbital elements for the entire evolution of the primary using the stellar data from MESA, but with a time step dictated by Runge-Kutta integrator. We stop the integration if the primary star fills its Roche lobe. The Roche radius is calculated using the approximation of \citet{Eggleton1983} for circular orbits.

\subsection{Rotation of the Main Sequence Progenitor of the Primary}
\label{sec:Rotation}

Quite independently from the changes in the primary's angular momentum at the hand of tides and mass loss, the star also changes its spin frequency, $\Omega$, due to {\it conservation} of angular momentum under an evolving radius. To calculate the initial value of the angular momentum of the primary star, we use initial surface velocities, $V_0$, from \citet[][]{Ekstrom2012}, who adopted zero-age main sequence (ZAMS) rotation values to be 40 per cent of the break-up velocity of the star, and we calculate the initial moment of inertia according to the formula $I\,=\,0.1\,(M_{{\rm env}}\,R^2) + 0.21\,(M_{\rm c}\,R_c^2)$ \citep{Hurley2000}, where the symbols have their usual meaning, and $M_{\rm c}$ and $R_{\rm c}$ are the primary star's core mass and core radius, respectively. Once the moment of inertia is calculated, we use the evolution of the stellar angular momentum from Equation~2 to calculate the surface rotation of the star during the entire evolution. 
 
Calculating the moment of inertia with this fitting formula results in values that are very close to what is calculated by direct integration of the stellar structure, with excursions of up to 20 per cent only during AGB pulses. In Section~\ref{sec:maximumengulfment} we will compare the orbital elements evolution using the moment of inertia determined by the fitting formula here, with the evolution obtained when calculating the moment of inertia by integrating the stellar structure as a function of time. 

The main sequence surface rotation rates of \citet{Ekstrom2012} are substantially larger than those calculated using equation 14 of \citet{Hurley2002} for lower mass stars ($<$2.5~\msun; Table~\ref{tab:Velocitytable}), but reasonably close at higher mass. 
We insured that the value selected did not affect our results substantially (namely the values of the capture distance discussed in Section~\ref{sec:maximumengulfment}) by running identical models with a range of initial rotation velocities. For example, for the 1.5~\msun\ star, we used initial surface velocity values of 50 and 250~km~s$^{-1}$, in addition to the value of 150~km~s$^{-1}$. The difference in capture distance is below 1 per cent for all companions.

As explained in Section~\ref{sec:mesa}, the stellar evolutionary models were calculated with no rotation. Had we calculated the models with rotation we would have obtained different values of stellar radius, mass and stellar spin as a function of time. All these values  are used by the tidal equations. However, for typical single star rotations, and in particular during the only important evolutionary phases as far as tides are concerned, mass and radius are not particularly different with or without rotation. 

The value of the stellar spin, $\Omega$, also used by the tidal equations, deserves a comment. Tides alter the value of $\Omega$  compared to the single star value, something that MESA in single star mode, even if run with rotation, does not account for. The value of $\Omega$ is therefore calculated by the tidal equations. The most recent versions of MESA allow one to run the stellar model under the influence of a companion, including a tidal prescription that uses the same basic theory as used here. This is likely more correct, but also more time consuming. For example, recently, \citet{GarciaSegura2016} carried out such a calculation using a single 2.5~\msun\ star showing, among other things, that the initial value of the stellar spin velocity is not influential on the outcome of the tidal interaction.  

\begin{table}\centering
\begin{centering}
\begin{tabular}{ccc}
\hline
\hline
 Mass & Velocity & Velocity\\
  (M$_\odot$)        & (km s$^{-1}$)& (km s$^{-1}$)\\
   &  Ekstrom   &  BSE  \\
 \hline
 \hline
0.8 & 40 & 10 \\
0.9 & 50 & 15\\
1.0 & 50 & 20  \\
1.1 & 50 & 30\\
1.2 & 85 & 35 \\
1.25 & 100 & 45\\
1.35 & 100 & 50   \\
1.5 & 150 & 65\\
1.7 & 176 & 90 \\
1.8 & 180 & 100\\
2.0 & 185 & 125   \\
2.5 & 190 & 175\\
3.0 & 195 & 210   \\
\hline
\hline
\end{tabular}
\caption{Surface rotation velocity of the star at the ZAMS using the values of \protect\cite{Ekstrom2012} and of \citet[][BSE]{Hurley2002}.}
\label{tab:Velocitytable}
\end{centering}
\end{table}

\section{The evolution of the primary's stellar Parameters}
\label{sec:mesa}
\begin{figure*}
\includegraphics[height=210mm]{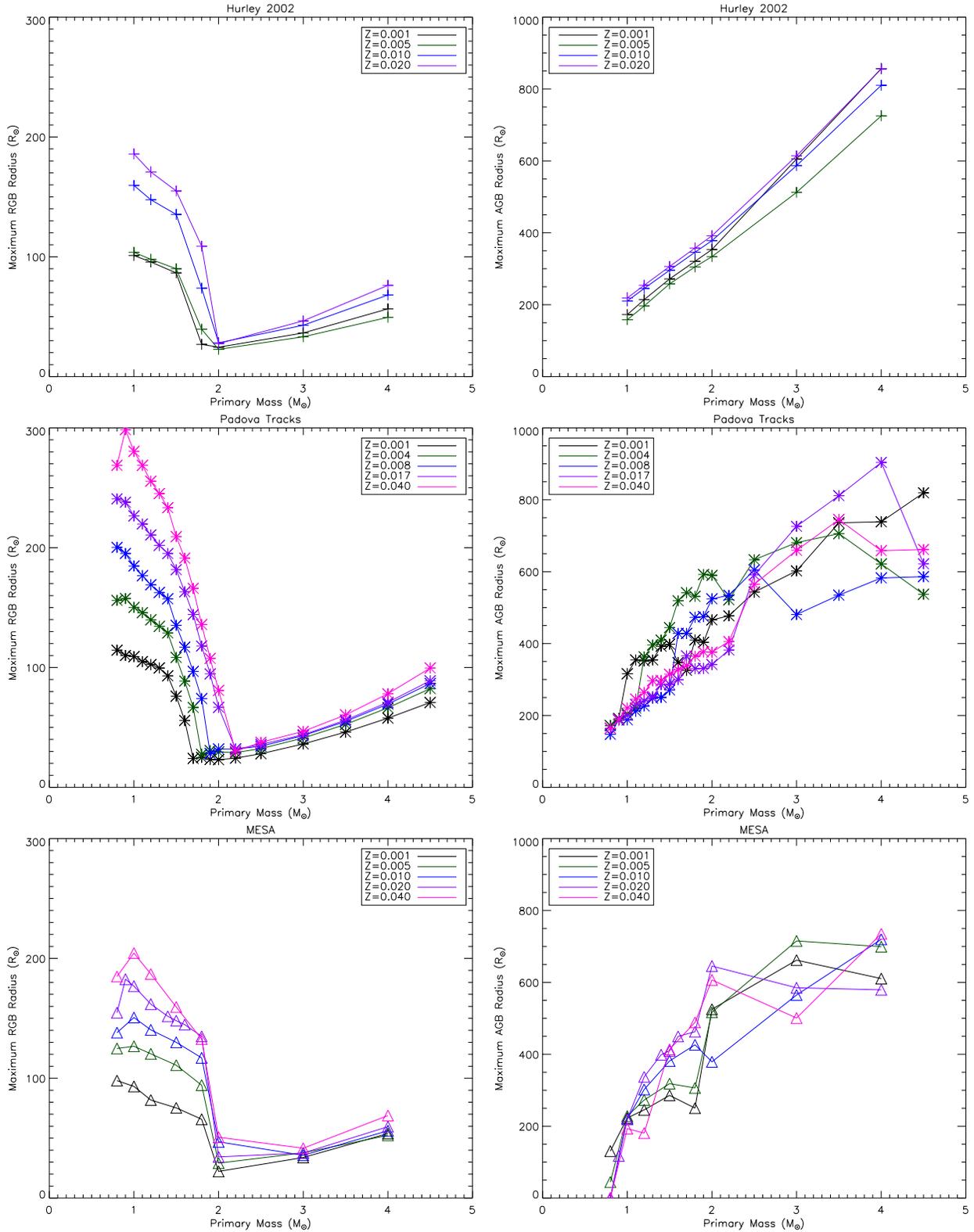}
\caption{ The comparison of maximum RGB (left) and AGB (right; if the model departs the AGB in the inter-pulse phase, the the maximum AGB radius is the radius of the last thermal pulse) stellar radii as a function of ZAMS mass for three different stellar evolutionary models. The MESA models for 3 and 4 M$_\odot$ with Z=0.02, and for 2, 3 and 4 M$_\odot$ with Z=0.04 do not reach the end of the AGB, so the maximum AGB radius plotted is that of the last model before convergence issues arise. The actual radii however should not be significantly larger than those we have plotted (see Section \ref{sec:mesa}). \label{fig:MultiRadcom}}
\end{figure*}  

In this study we use the 1D stellar evolution code MESA (Version 4219; \citealt{Paxton2011, Paxton2013}) to calculate a series of stellar evolutionary models for single, non-rotating stars with ZAMS mass between 0.8 and 4.0~\msun, with metallicities $Z = 0.001, 0.005, 0.01, 0.02$ and $0.04$. The opacity tables adopted are those detailed in section~4.3 of \citet{Paxton2011}. The MESA mass-loss rate prescriptions are those of \cite{Reimers1975} and \cite{Bloecker1995} for the RGB and the AGB phases, respectively. The mass-loss rates depend on the mass-loss coefficients $\eta_R$ and $\eta_B$, respectively. The Reimer's mass-loss coefficient, $\eta_R$ is fixed at 0.5, which is consistent with the value used by \cite{Passy2012}, \cite{Ekstrom2012} and the recently calculated value of 0.48 \citep{McDonald2015}. The $\eta_B$ coefficient ranges between 0.02 \citep{Ventura2013, Ventura2012} and  0.10 \citep{Herwig2004}. The coefficient $\eta_B$ used in this work is 0.10 for stars with ZAMS mass $\leq$ 2~M$_\odot$, and 0.05 for stars with masses $\gtreq2$~M$_\odot$ \citep[see] []{Paxton2011}. These coefficients are not calculated {\it ab initio}, but are instead calibrated using observations of the tip of the RGB or the luminosities of post-AGB stars in clusters. The effect of doubling the coefficient from 0.1 to 0.2 is to decrease the maximum AGB radius by $\sim$10 per cent, while halving it to 0.05 only increases the radius by $\sim$1 per cent.
       
\begin{figure*}
\includegraphics[height=65mm]{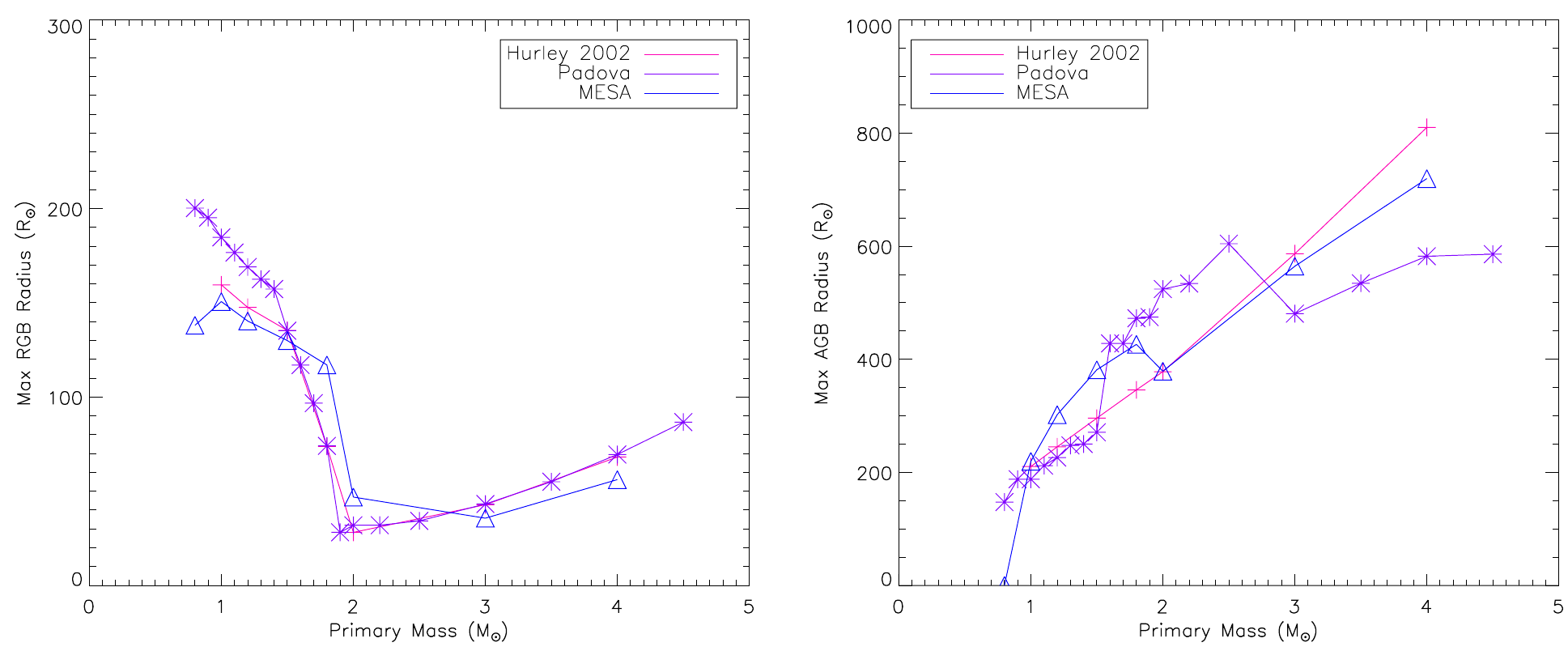}\label{fig:MultiMetallicity}
\caption{Comparison of maximum RGB (left) and AGB (right) stellar radii as a function of main sequence mass for three different stellar evolution models. The MESA (triangles) and the SSE (plusses) models of \protect\cite{Hurley2000} were calculated with Z=0.01, while the Padova models of \citet[][asterisks]{Bertelli2008,Bertelli2009} are for Z=0.008.}
\label{fig:MultiMetallicity}
\end{figure*}  

We have used the ``simple" atmosphere option for the star's outer boundary, where the atmosphere limit is defined at the point where the optical depth $\tau$ is 2/3 \citep[][eq.~3]{Paxton2011}. However, neither \citet{Passy2012b} nor \citet{Stancliffe2016} found any significant differences in the radii of the stars using the default or other atmosphere calculation options. We have run a 1.5~\msun\ stellar structure from the main sequence to the white dwarf regime using the single atmosphere and the Eddington grey integration \citep[][section~5.3]{Paxton2011} options, and found negligible changes. This said, it is likely that for AGB stars the value of the radius defined by optical depth properties of the very extended atmospheres is not the best representation of the size of the star. 

In Fig.~\ref{fig:MultiRadcom} we present the maximum radii on the RGB (left column) and AGB (right column) for  three stellar codes,  the single star evolution (SSE) code \citep{Hurley2000}, the Padova stellar evolution code\footnote{http://stev.oapd.inaf.it/YZVAR/; more recent models from the Padova group have been presented, e.g., by \citet{Bressan2012} and \citet{Marigo2013}.} \citep[][]{Bertelli2008, Bertelli2009} and the MESA code used here \citep[][]{Paxton2011, Paxton2013}. The  models of \cite{Hurley2000} can be calculated for metallicities Z = 0.0001 - 0.03  and ZAMS masses in the range 0.5 - 50 M$_\odot$, but use fitting formulae for each evolutionary stage. One major advantage of SSE is its ability to calculate stellar evolutionary models in less than a second, which makes it able to be included directly in population synthesis codes. 

In Fig.~\ref{fig:MultiMetallicity} we carry the comparison between different stellar models on the same plot, for a metallicity Z=0.01. In Fig. \ref{fig:Radiusratio} we plot the ratio of maximum AGB to maximum RGB radii as a function of primary mass for the the MESA models and as calculated by \citet{Soker1998}, using stellar evolutionary models of \citet{Iben1985}. From these three figures we draw the following conclusions.

The main difference between the modelling codes compared here is seen for lower mass stars on the RGB, particularly at lower metallicities, where the Padova models of \citet{Bertelli2008} predict substantially larger maximum RGB radii than either MESA or the  fitting formulae. Using the Padova models would therefore result in a larger number of predicted RGB interactions for these stars. The  models use fits to the stellar evolutionary calculations of \cite{Pols1998}, which arguably do not take into account the most recent updates in stellar astrophysics like the changes in opacity tables, e.g., \cite{Ferguson2005}, or the latest prescriptions for  convective overshooting \citep{Herwig2000}. The Padova stellar evolution models of \citet{Bertelli2008} ignored mass loss during the RGB. A star between 0.8 and $\sim\,$1.8~M$_\odot$, loses between 0.09 and 0.23~M$_\odot$ during the RGB. Ignoring the RGB mass loss, results in a star with higher envelope mass and a change in the evolution of the stellar radius. The issue for stars with mass 3 to 4 M$_\odot$ is not as pressing, because the mass lost during the RGB is less than 1 per cent of the initial mass. Hence, the peak RGB radii obtained for these stars do not change significantly with the inclusion or omission of mass-loss during the RGB. Another possible drawback of the Padova models of \citet{Bertelli2008}, when used to predict tidal effects is that they use a mean radius during the AGB and do not resolve the peak and trough of each thermal pulse. 

Note as well that, as discussed by \citet{Villaver2014} the extent of the RGB radius is very sensitive to the metallicity,  especially around the transition mass that marks the boundary between degenerate and non-degenerate cores. Models that develop electron degenerate helium cores after the end of central hydrogen burning have an extended and luminous RGB phase prior to core helium ignition and therefore represent the most interesting arena for tidal capture on the RGB. Thus the mass-loss prescription in the low mass stellar range strongly influences the tidal capture distance \citep{Villaver2014}. The mass-loss parameter adopted in this paper for the RGB results in a conservative estimate to the tidal capture distance. 

The BSE maximum AGB radii increase monotonically with increasing initial mass, due to the use of fitting formulae that ignore the thermal pulses. On the other hand, both the Padova models of \citet{Bertelli2008} and our own display an irregular behaviour of maximum AGB radii. Overall the radii compare well across different calculations, but there is scatter and a clearly complex behaviour. This is due to a range of reasons. First of all, if the AGB evolution ends just before the next thermal pulse, something that can be caused by very minor differences in the previous evolution, the maximum AGB radius will be that of the last thermal pulse, or 10-20 per cent lower than if the next thermal pulse had happened (see for example Fig.~\ref{fig:Avst2}, lower panel). Additionally, during this phase of stellar evolution, complex processes are involved that are fraught with uncertainty, such as the efficiency of the third dredge-up. Convection and dredge-up are non-local, three dimensional processes  and thus an accurate determination of thermal pulses should use a three-dimensional stellar evolution code (see \citealt{Pols2001} for more details). In addition, upper AGB stellar envelopes are almost unbound. Changes during this phase are dynamical and small changes in any one quantity, such as the shell burning rate, have large effects on all other structural quantities. In practice large oscillations can become unphysical and preclude convergence. A way to circumvent this problem and continue to evolve a star past the AGB phase, is to increase artificially the mass-loss rate on the upper AGB so as to terminate this phase before the model becomes unstable. We could not converge MESA models for  AGB  stars with masses of 3 and 4~M$_\odot$ and Z= 0.02, nor for masses of 2, 3 and 4~M$_\odot$ for Z= 0.04. This is likely due to larger envelope opacities leading to more extreme envelope response to changes, particularly for the more massive stars in our range. For these models we therefore plotted in Figs.~\ref{fig:MultiRadcom} to \ref{fig:Radiusratio}, the AGB pulse radius value preceding the last converged model, noting that it is a lower limit. 

 The maximum RGB radius has  a minimum at $\sim$1.8~\msun\ (Figs.~1 and \ref{fig:MultiMetallicity}).  This ``kink" must have a profound effect on the population of PN that derives from binaries \citep{Soker1998}. From Fig.~\ref{fig:Radiusratio} it is clear that the lower mass stars ($M\lesssim2$~\msun, which are relatively more numerous), have a low AGB-to-RGB radius ratio, while more massive stars have a much larger ratio. The lower the ratio the larger the relative number of RGB to AGB interactions. For binaries to interact on the AGB and impact the PN they need to have orbital separations tuned to be large enough to avoid tidal capture on the RGB, but small enough to be captured on the AGB. For primary stars with a mass smaller than approximately 2~\msun\ the range of such separations is smaller. For example a 1-\msun\ star with Z=0.02 has a maximum RGB radius that is only just smaller than its maximum AGB one (180~\rsun\ vs. 230~\rsun). The exact limits of this range depend on the particular stellar evolutionary model used. We discuss the consequences of this range in Section~\ref{section:keyhole}.

Finally there are a host of additional effects that are not considered by stellar structure and evolution calculations.  One of them is the fact that the giant star will be substantially distorted by the presence of an orbiting companion, particularly for large mass ratios. Such a star likely presents properties that could substantially diverge from those calculated here and that could affect the the evolution as well as the tidal behaviour of the star. One such effect could be on the mass-loss of the star. \citet{Dijkstra2006} determined, for example, that the small equatorial bulge that develops in rotating giant stars would substantially enhance the formation of dust, which may in turn result in enhanced mass-loss. In the far larger tidal bulges induced by the tidal interaction between a giant and a nearby massive companion this could be of great consequence. However, we cannot at present consider these effects.  Fortunately the binaries of greatest interest to this paper, those that result in CE binary central stars of PN, are those of lower mass with relatively low mass companions \citep[see, e.g.,][table~2]{DeMarco2009}, where this effect might not be as prominent.
 \begin{figure}
\includegraphics[width=85mm]{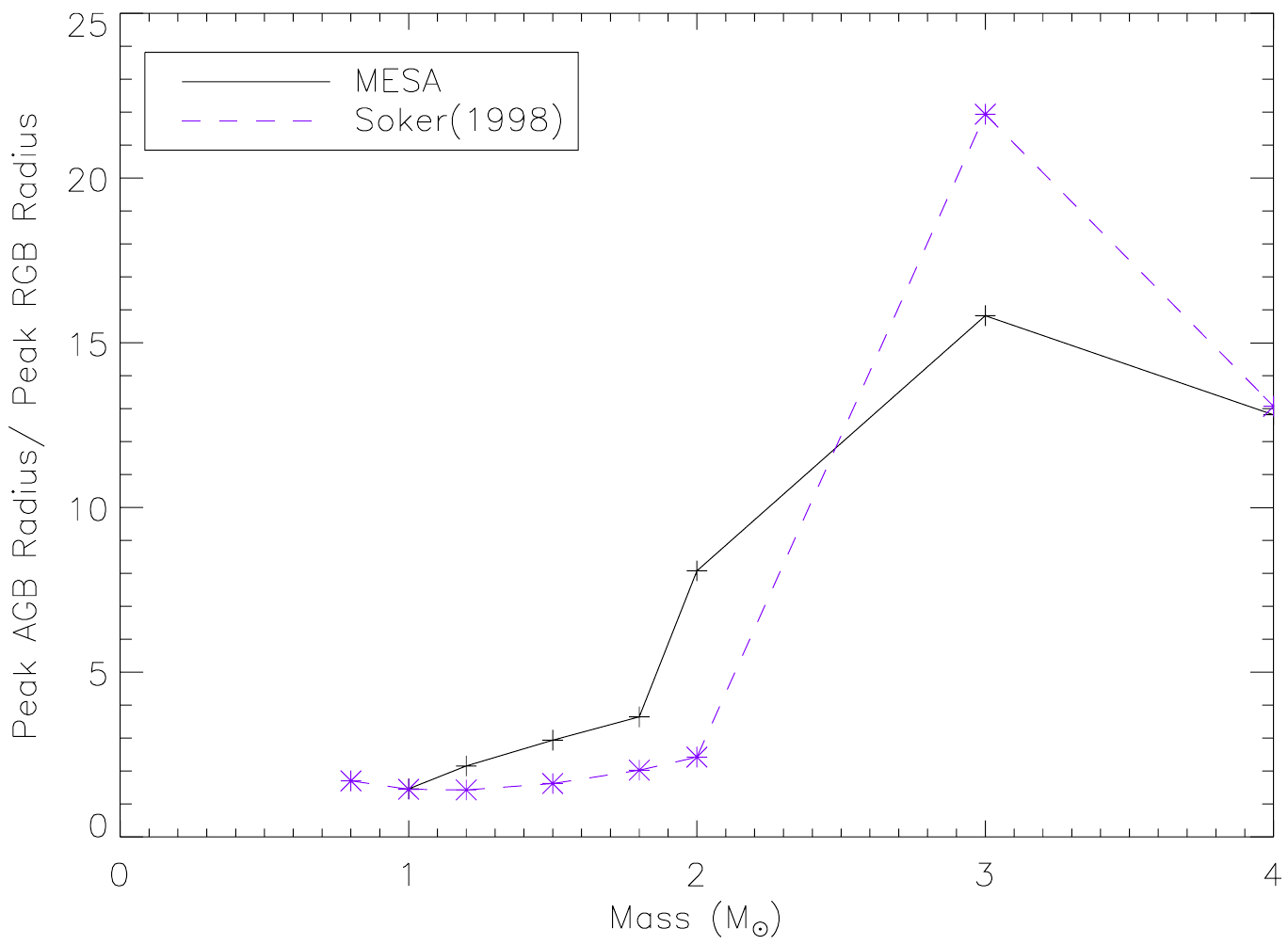}
\caption{A comparison of the ratio between the peak AGB and peak RGB radii as a function of ZAMS mass for MESA stellar models (Z=0.01), and using the prescription of  \protect\cite{Soker1998}.}
\label{fig:Radiusratio} 
\end{figure}

\section{The Maximum Capture Distance as a Function of Primary and Companion Mass}
\label{sec:maximumengulfment}

Here we determine the maximum orbital extent for which companions will be tidally captured into a Roche lobe overflow interaction with RGB and AGB stars. We have nicknamed this orbital separation the {\it maximum capture distance} or MCD. We carry out the integration for primary masses of 0.8, 1.0, 1.2, 1.5, 1.8, 2.0, 3.0 and 4.0~\msun, evolved using a metallicity $Z=0.01$ to circumvent the lack of model convergence problems at larger metallicities explained in Section~\ref{sec:mesa}. We use approximately 50 companions with masses chosen between 1~\mjup\ and a mass ratio $q=M_2/M\lesssim 1$. We start with describing the behaviour of 1~\mjup\ and 0.15~\msun\ companions orbiting a 1.5~\msun\ star to emphasise the differences brought about by stronger tides and spin-orbit interaction.

\subsection{The tidal interaction between a 1-M${\rm{_J}}$ Companion and its 1.5-M$_\odot$ primary star}
\label{sec:Planet}

\begin{figure}
\includegraphics[width=92mm]{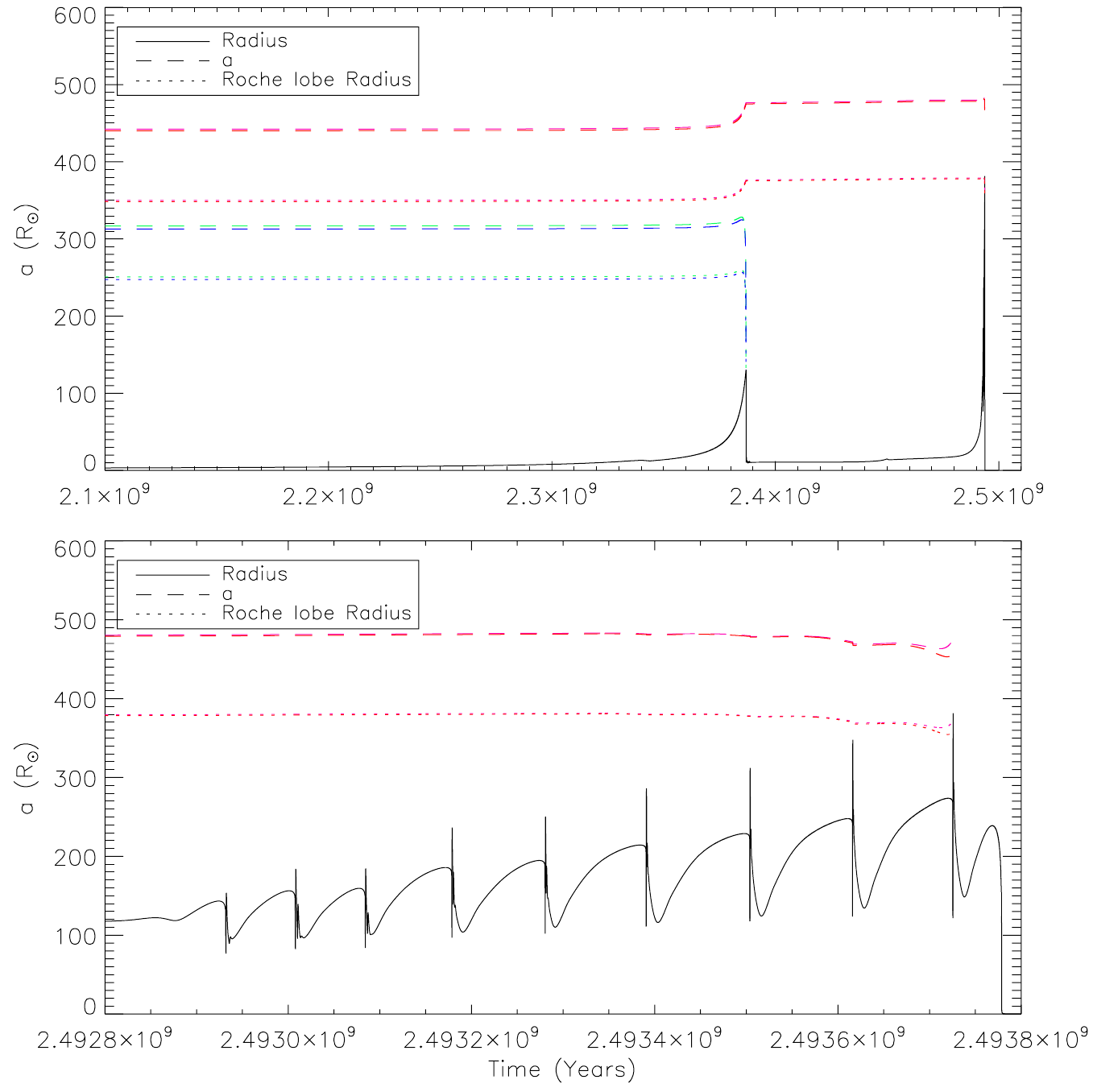}
\caption{Top: Temporal evolution of the stellar radius (solid line), semi-major axes (dashed lines) and Roche-lobe radius (dotted lines) during the entire stellar evolution for a system comprising of a 1.5~M$_\odot$ primary (Z=0.01) and a 1~M$_{\rm{J}}$ companion at initial separations of 310~$R_\odot$ (blue and green lines) and 440~$R_\odot$ (red and pink lines). Bottom: close-up of the evolution as the star evolves through the AGB thermal pulses for the system with an initial semi-major axis of 440~R$_\odot$. { The red and blue lines used the approximate prescription for the moment of inertia calculation detailed in Section~\ref{sec:Rotation}, while pink and green lines used an accurate calculation based on the stellar structure.}}
\label{fig:Avst2}
\end{figure} 

The top panel of Fig. \ref{fig:Avst2} shows  the tidal evolution of a binary system with a 1.5~M$_\odot$ (Z=0.01) and a 1~M$_{\rm{J}}$ companion. The MCD on the RGB is 310~\rsun\footnote{We round these values to the nearest 5~\rsun. However, the actual uncertainty may well be higher than such a value and indeed vary for different cases.} ($\sim$1.4 au or 2.4 times the maximum RGB radius of $\sim$130~\rsun), while for a capture on the AGB the separation is 440~R$_\odot$ ($\sim$2 au, 1.2 times the maximum pulse AGB radius of 390~\rsun, or 1.6 times the maximum inter-pulse radius of 270~\rsun). Thus, the population of binaries with a 1.5~\msun\ primary with initial orbital separations between 310 and 440~R$_\odot$ are captured during the AGB. This shows that the the range of initial orbital separations that lead to an interaction on the AGB discussed in Section~\ref{sec:mesa}, is made relatively smaller by tides.

{ In Fig.~\ref{fig:Avst2} we also show how the MCD varies if we use a moment of inertia calculated using fitting formulae (Section~\ref{sec:Rotation}) compared to one integrating the stellar structure: the difference between the two MCD values is 1 per cent.}

As the star evolves through the RGB, it loses $\sim$0.12~M$_\odot$. This mass loss causes an increase in the semi-major axis (Equation \ref{eq:newaa1}) of $\sim$40~R$_\odot$ (see \citet{Villaver2014} for a discussion of the influence of the adopted parameterisation of the RGB mass-loss on the orbital evolution). The tidal interaction can also increase the semi-major axis if $(1\,-\,\Omega/\omega)$ is negative in Equation \ref{eq:newaa1}. We will discuss this effect further in Section \ref{sec:Star}.

The bottom panel of Fig. \ref{fig:Avst2} shows a close up of the capture of the companion during the AGB thermal pulses, when the system is subject to the effect of strong tides as well as strong mass loss. Here we observe that the strongest tidal interaction (shortening of the orbital separation) happens at the peak of the {\it inter-pulse} radius. The pulse radius is much larger, but lasts but a short time ($\sim$100 to 1000 years), making the tide relatively ineffective. However, if the companion finds itself within reach of the pulse radius (in other words the primary fills its Roche radius during the pulse) then the capture will take place during the pulse. This is the case in Fig.~\ref{fig:Avst2} where the companion is captured during the last pulse. To demonstrate further the role of the thermal pulses on the MCD, we plot, in Fig.~\ref{fig:Capturepulse}, the evolution of the semi-major axis for initial values between 415 and 450~R$_\odot$ in increments of 5~\rsun.
 \begin{figure}
 \includegraphics[width=89mm]{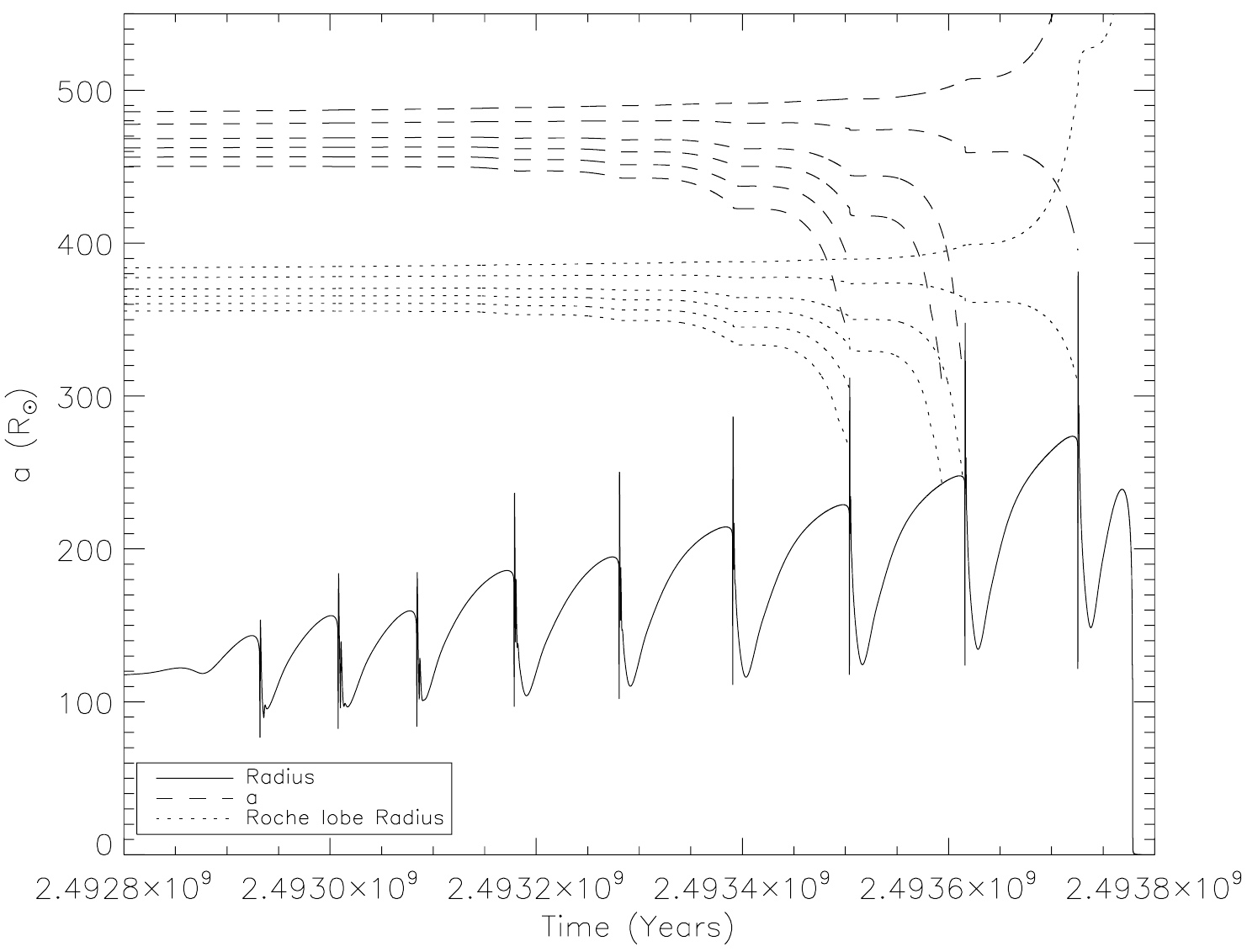}
 \caption{Temporal evolution of the stellar radius (solid line), semi-major axis (dashed line) and Roche-lobe radius (dotted line) during the AGB thermal pulses for systems comprising of a 1.5~M$_\odot$ primary, a 1~M$_{\rm{J}}$ companion and a range of initial separations. \label{fig:Capturepulse}}
 \end{figure}

From this figure it is clear that pulse captures only happen when the stellar radius grows  to meet the Roche-lobe radius. Tidal action predominately happens during the pre-pulse time. Combining our work with that of \citet[][their figure 2]{Mustill2012}, who investigated Earth and Jupiter mass companions, we see that the weaker the tide (the lighter the companion) the more companions are captured during the pulses; the stronger the tide, the more companions are captured during the pre-pulse time. This indicates that in population synthesis calculations the prescription for the MCD for an AGB interaction should be based on a careful assessment of pulse and inter-pulse radius for each mass regime.

Finally, we point out that \citet{Privitera2016} showed that models including rotating stars make the MCD for planets smaller by 20 per cent, compared to non-rotating ones. This would allow more planetary companions to survive the RGB and be available for an interaction on the AGB.

\subsection{The tidal interaction between a 0.15-M$_\odot$ Companion and its 1.5-M$_\odot$ Primary Star and the Effects of Spin-Orbit Interaction} 
\label{sec:Star}

As the companion mass is increased the tide will be stronger and we might expect a tidal capture to happen even for companions with larger initial separations (see Equations \ref{eq:newaa1} and \ref{eq:newOmega1}). However, we also expect that a more massive companion will more easily synchronise its orbit with the spin of the primary, thus halting the orbital decrease. In this section we examine the orbital evolution of a close binary, subject to these two competing phenomena, by increasing the mass of the companion to 0.15~M$_\odot$. This is the mass of an M5V star \citep[see their table C1]{DeMarco2013}, a common type of companion to central stars of PN. Spin-orbit interaction will transfer orbital angular momentum to the star. This will contribute to spinning up the giant, something that can lead to an axi-symmetric PN  \citep{Nordhaus2006,GarciaSegura2014}.

\begin{figure}
\includegraphics[width=90mm]{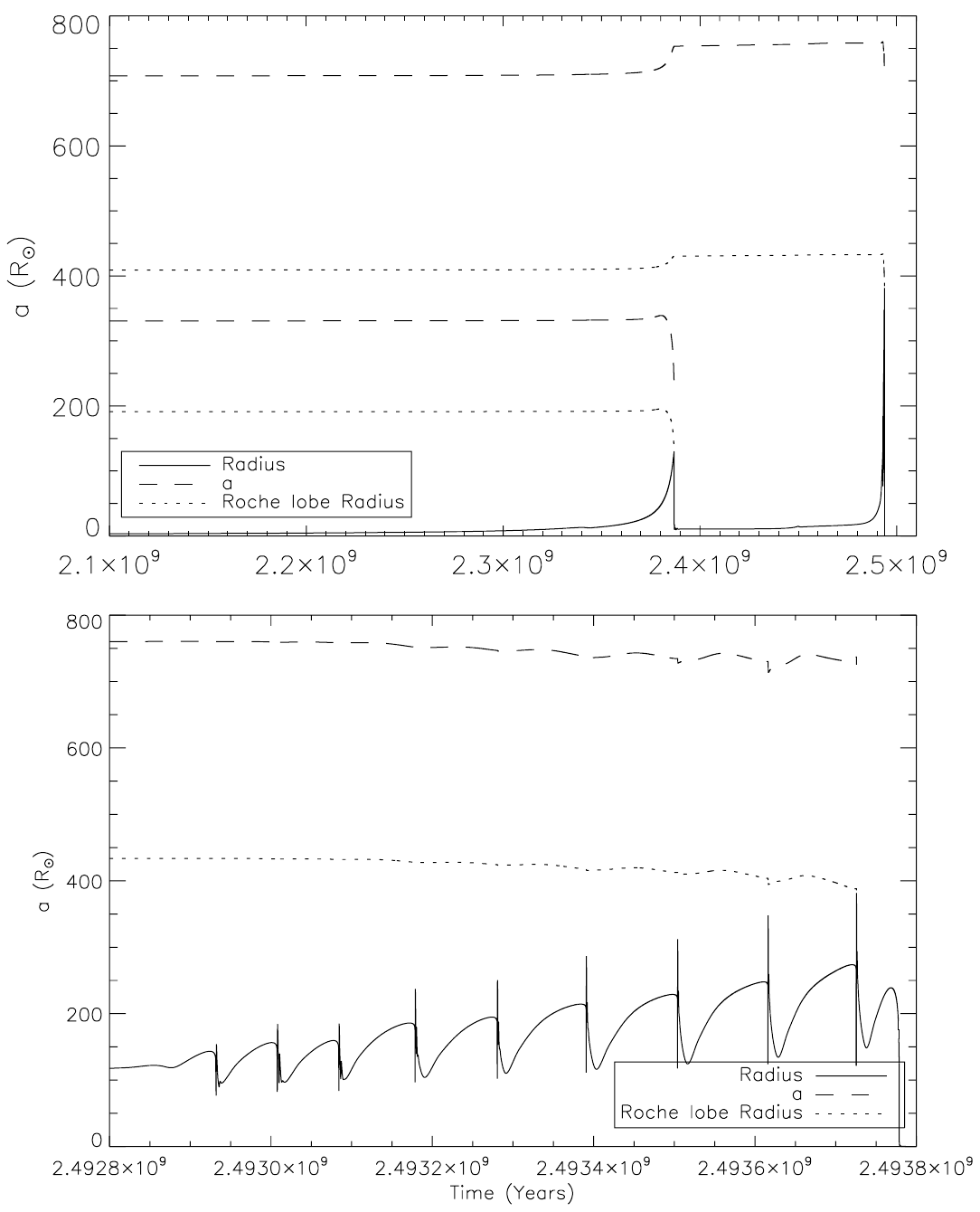}
\caption{Top: the evolution of the semi-major axis, $a$, for a 1.5~\msun\ primary and a 0.15 $M_\odot$ companion with initial semi-major axes of 340 and 720~R$_\odot$. Bottom:  close-up of the evolution in semi-major axis during the AGB thermal pulses. Solid line: stellar radius; dashed line: semi-major axis; dotted line: Roche lobe radius.}
\label{fig:Capture150}
\end{figure}
 
Fig. \ref{fig:Capture150} depicts the evolution for two values of the initial semi-major axis of our system. The MCDs are 340~R$_\odot$ and 720~R$_\odot$ for the RGB and AGB, respectively or 2.6 and 1.8 times their respective maximum RGB and {\it pulse} AGB radii (2.7 times the maximum AGB inter-pulse radius). The MCDs for a 1-M${\rm{_J}}$ companion were 310 and 440~R$_\odot$, respectively (Section \ref{sec:Planet}) or 2.4 and 1.2 times the maximum RGB and {\it pulse} AGB radii (1.6 times the maximum AGB inter-pulse radius). This indicates that as the companion mass increases, the MCD increases, due essentially to stronger tides. However, the increase would have been larger had we not included spin orbit-interaction, which reduces the strength of the tide by spinning up the giant star, as we are about to explain. 
Also, the $q\,(1\,+\,q)$ factor in Equation \ref{eq:newaa1} is larger during the AGB than during the RGB, because the  mass loss from the primary during the AGB reduces the mass of the primary. The companion also accretes some mass, although in this case accretion only increased the companion mass by $\sim\,2$ per cent.

 \begin{figure}
\includegraphics[width=90mm]{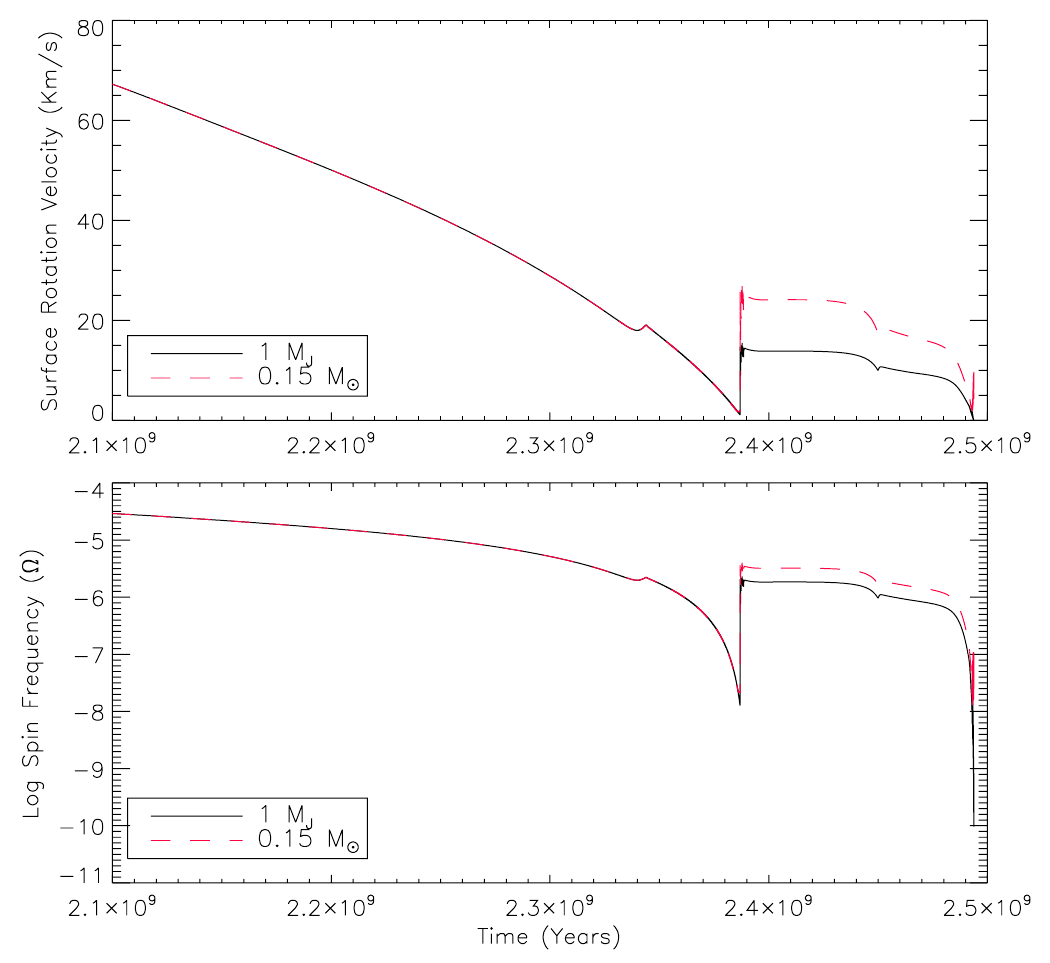}
\caption{The evolution in the surface velocity (top panel) and the spin frequency (bottom panel) for a primary of 1.5 M$_\odot$ with binary companions of 1 M${\rm{_J}}$ with an initial separation of 340~\rsun\ (black solid line) and 0.15 M$_\odot$, with an initial separation of 720~\rsun (red dashed line).  \label{fig:Spin_comparison}} 
\end{figure}
 
Fig. \ref{fig:Spin_comparison} shows the comparison of spin frequencies and surface rotation velocities of the primary for the 1~M${\rm{_J}}$ and 0.15~M$_\odot$ companion cases with initial semi-major axes of 440 and 720 R$_\odot$, respectively. The heavier companion causes a stronger tide causing a larger amount of orbital angular momentum to be transferred to the primary. During the RGB, the heavier companion spins up the primary far more than the lighter companion. At the tip of the RGB the values are 1~km~s$^{-1}$ and 2~km~s$^{-1}$ for the 1~M${\rm{_J}}$ and 0.15~M$_\odot$ companions, respectively. Neither companion is engulfed by the primary at this stage and the systems continue to evolve as binaries. During the horizontal branch evolution, the surface rotation velocity of the star increases due to radial contraction and is considerably larger for the system with the heavier binary companion, 29~km~s$^{-1}$ vs. 15~km~s$^{-1}$, for the system with the 1-M${\rm{_J}}$ companion.
 
 \begin{figure}
 \includegraphics[width=90mm]{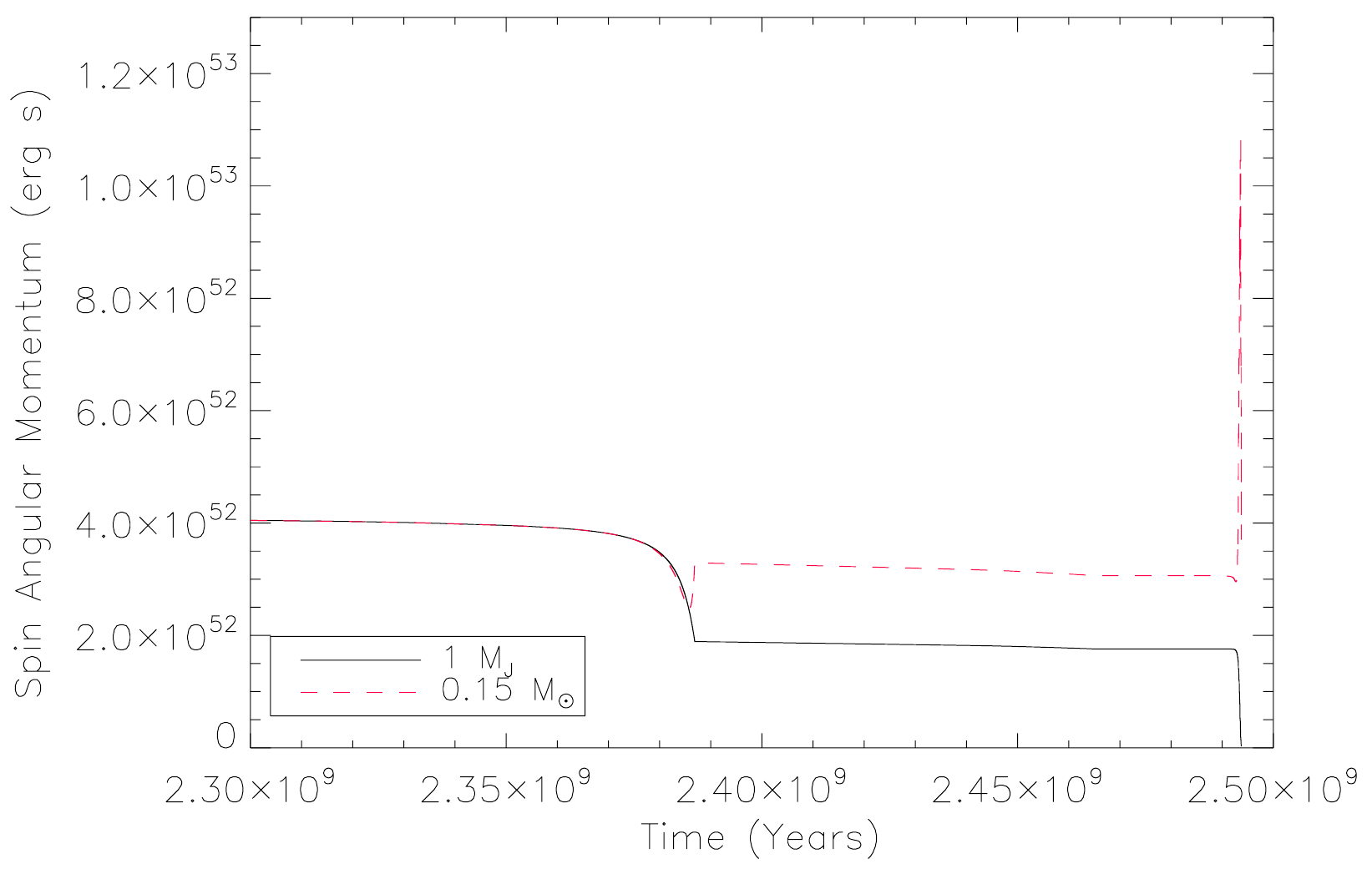}
 \caption{The spin angular momentum of a 1.5-\msun\ primary orbited by companions of 1 M${\rm{_J}}$ (black solid line) and 0.15 M$_\odot$ (red dashed line). The spin angular momentum of the primary remains approximately constant during the main sequence, then it decreases due to mass loss during the RGB. The companion with a higher mass has the ability to spin up the primary during the RGB and the AGB.\label{fig:Mom_com}}
 \end{figure} 
        
Another way to visualise the effect of the spin-orbit interaction is shown in Fig.~\ref{fig:Mom_com}, depicting the change in spin angular momentum of the primary in the same two cases as Fig.~\ref{fig:Spin_comparison}. The angular momentum of the primary starts to decrease as the star looses mass during the RGB. However, for the binary system with a more massive companion, the initial decrease in spin angular momentum is followed by an increase due to angular momentum being transferred from the orbit. A system with a 1-M${\rm{_J}}$ companion is unable to transfer orbital angular momentum to the primary, because the tidal coupling is weaker. During the horizontal branch and the early AGB phases the primary with a heavier companion transfers a small fraction of its angular momentum back to the orbit. However, later on, when the primary's radius increases further on the AGB, the strong tide dominates and the star's angular momentum increases: while some angular momentum is lost because of mass loss, more angular momentum is transferred to the star from the orbit. 

This reinforces and completes the conclusions of \citet{Nordhaus2006} and \citet{GarciaSegura2014}: these authors argue that single AGB stars cannot sustain a global magnetic field, necessary for shaping the PN and that a companion interacting with the AGB star is needed in order to supply angular momentum. Here we see that the interaction that donates angular momentum to the star has a tidal component, starting when the companion is almost two stellar radii distant, and later on, a direct interaction component, as the primary fills its Roche lobe and potentially enters a CE interaction phase. 
 
\begin{figure*}
\includegraphics[width=120mm]{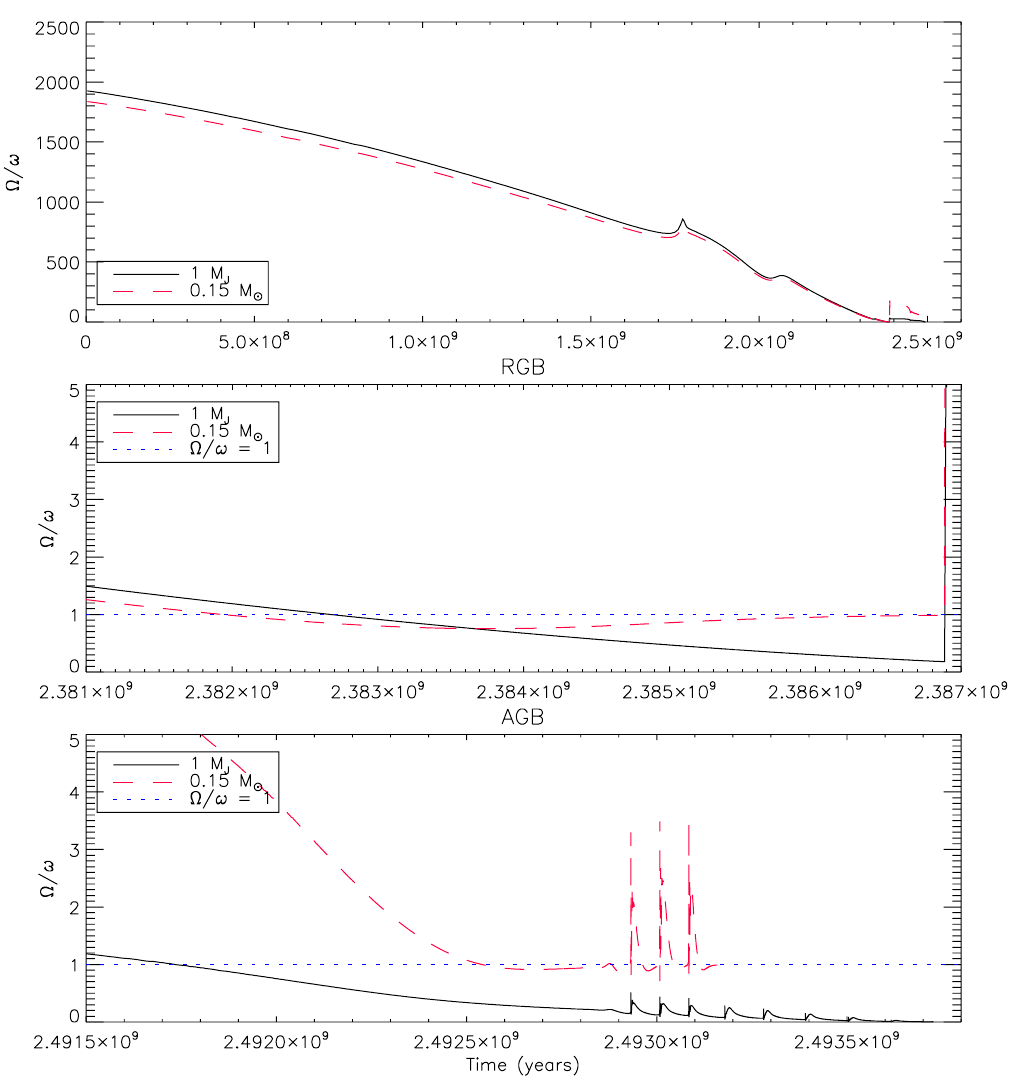}
\caption{Evolution of the synchronisation factor as the 1.5~\msun\ star evolves (top panel) for a 1-M${\rm{_J}}$ companion (black solid line) and 0.15-M$_\odot$ companion (red dashed line). The middle and bottom panels show details of the RGB and AGB evolution, respectively. The dotted line marks $\Omega / \omega$=1.\label{fig:Synchro_com}}
\end{figure*}   

Fig. \ref{fig:Synchro_com} shows the variation in $\Omega / \omega$, which we call the synchronisation factor, for the 1~M${\rm{_J}}$ and 0.15~M$_\odot$ companion cases. This time the two binaries were evolved with the same initial semi-major axis of 440 R$_\odot$. During the main sequence and the early RGB phases, the synchronisation factor is lower for a system with a heavier companion because the orbital frequency of the system is higher. The synchronisation factor is larger than unity, dictating an increase in the orbital separation, but the tide is very weak during this phase. During the RGB the synchronisation factors decrease to below unity, because the stellar spin decreases due to expansion. At this point the heavier of the two companions spins up the RGB star reversing the decreasing trend of the synchronisation factor and weakening the tidal interaction (Fig. \ref{fig:Synchro_com}, middle panel). 

Both systems have a synchronisation factor larger than unity during the horizontal branch, potentially dictating an increase in the orbital separation. However, the stellar radius during the horizontal branch is once again small so that the overall strength of the tide is too small for orbital evolution. During the horizontal branch, the heavier system has a synchronisation factor higher than the system with a lighter companion, due to the orbital angular momentum that was commuted into spin angular momentum during the preceding RGB phase. During the AGB the synchronisation factors plummet once again due to the slowing down of the expanding giant, but the system with a lighter companion attains a synchronisation factor smaller than unity before the heavier system. The lighter system maintains a synchronisation factor smaller than unity throughout the thermal pulses, which favours a capture. On the other hand, the system with a heavier companion also attains a synchronisation factor smaller than unity, but once again spin-orbit interaction increases the stellar spin and the factor rises. During the thermal pulses the factor oscillates below and above unity, due to conservation of spin angular momentum during the thermal pulses, until the companion is captured on the fourth thermal pulse.

While a strong synchronising effect must be taking place for the heavier companion during the pulsating AGB phase, and this must cause a lessening of the tide, the system with a heavier companion nonetheless achieves capture for a larger initial separation than the the system with the lighter  companion due to a larger value of $q$ in Eq.~\ref{eq:newaa1}. Also, a more massive companion results in a smaller Roche radius around the primary, which promotes an earlier capture. 

\begin{figure}
\includegraphics[width=85mm]{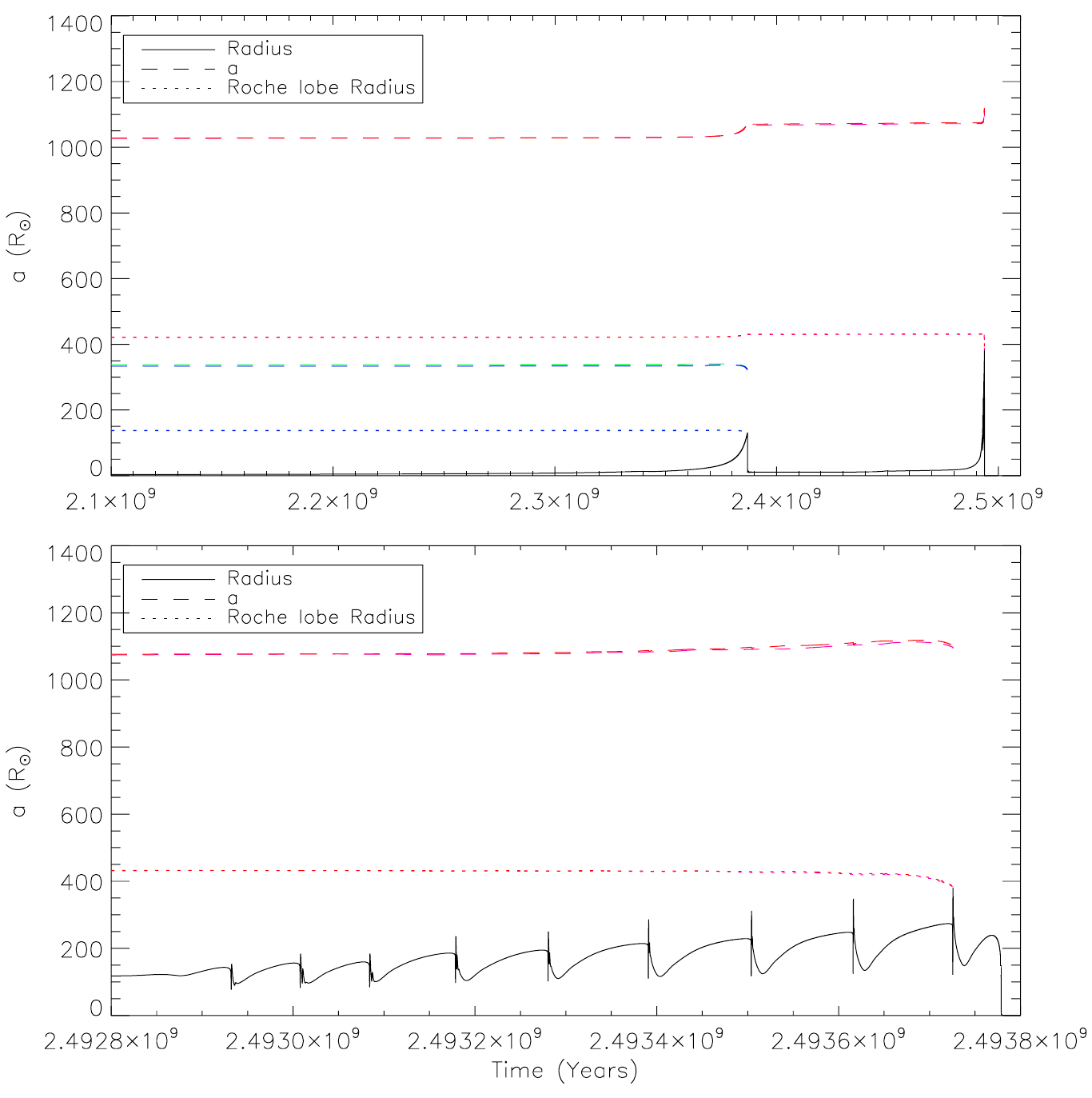}
\caption{Top: the evolution of the semi-major axis, $a$, for a 1.5~\msun\ primary and a 1.05 $M_\odot$ companion with initial semi-major axes of 340 and 1030~R$_\odot$. Bottom:  close-up of the evolution in semi-major axis during the AGB thermal pulses. Solid line: stellar radius; dashed line: semi-major axis; dotted line: Roche lobe radius.}
\label{fig:1MoCompanion}
\end{figure}  

For completeness we also show, in Fig.~\ref{fig:1MoCompanion}, the evolution of the orbital elements of the same primary, but with a much more massive, 1.05~\msun\ companion. The largest initial orbital separation that results in a capture on the RGB is 340~\rsun \footnote{This is the same value as for the 0.15-\msun\ companion. The reason is that while a 1.05-\msun\ companion excites a stronger tide, the same strong tide spins up the primary and reduces the effectiveness of the tidal action on the orbital separation.}, while for a capture on the AGB the MCD is 1030~\rsun. In this figure, as we have done in Fig.~\ref{fig:Avst2}, we also emphasise that calculating the tidal evolution using a moment of inertia calculated using the fitting formula shown in Section~\ref{sec:Rotation} is equivalent to using the more realistic integration of the stellar structure.

\subsection{The Effect of the Uncertainty on Giant Radii on the Maximum Capture Distance} 
\label{section:R_Uncert}

Before we discuss the MCD, we return to the issue of the large uncertainties on the AGB star radii discussed in Section~\ref{sec:mesa}.  Aside from difficulties in calculating the radii of upper AGB stars mentioned in Section~\ref{sec:mesa}, there are also uncertainties due to omitted physical processes, such as Mira pulsations. The strong dependence of the tidal orbital reduction (Eq.~1) on radius makes one wonder what the effect of the radius uncertainty could be. This is not a simple linear relation because radial changes influence the stellar spin rate, which impacts both the orbital separation change (Eq.~1) as well as the spin-orbit coupling (Eq.~2).

To determine how uncertainties in the radius propagate on the MCD, we have artificially increased the stellar radius of a 1.5~\msun\ stellar model by 10, 20, 30 and 40 percentage points along its evolution. In Fig.~\ref{fig:R_Uncert} we show the MCD as a function of companion mass for the unaltered, 1.5~\msun\ stellar evolutionary track, alongside each of these artificially-altered stellar evolutionary tracks. The increase in MCD is approximately proportional to the increase in maximum AGB radius, as can be seen in the bottom panel of Fig.~\ref{fig:R_Uncert}. 

\begin{figure}
\includegraphics[width=87mm]{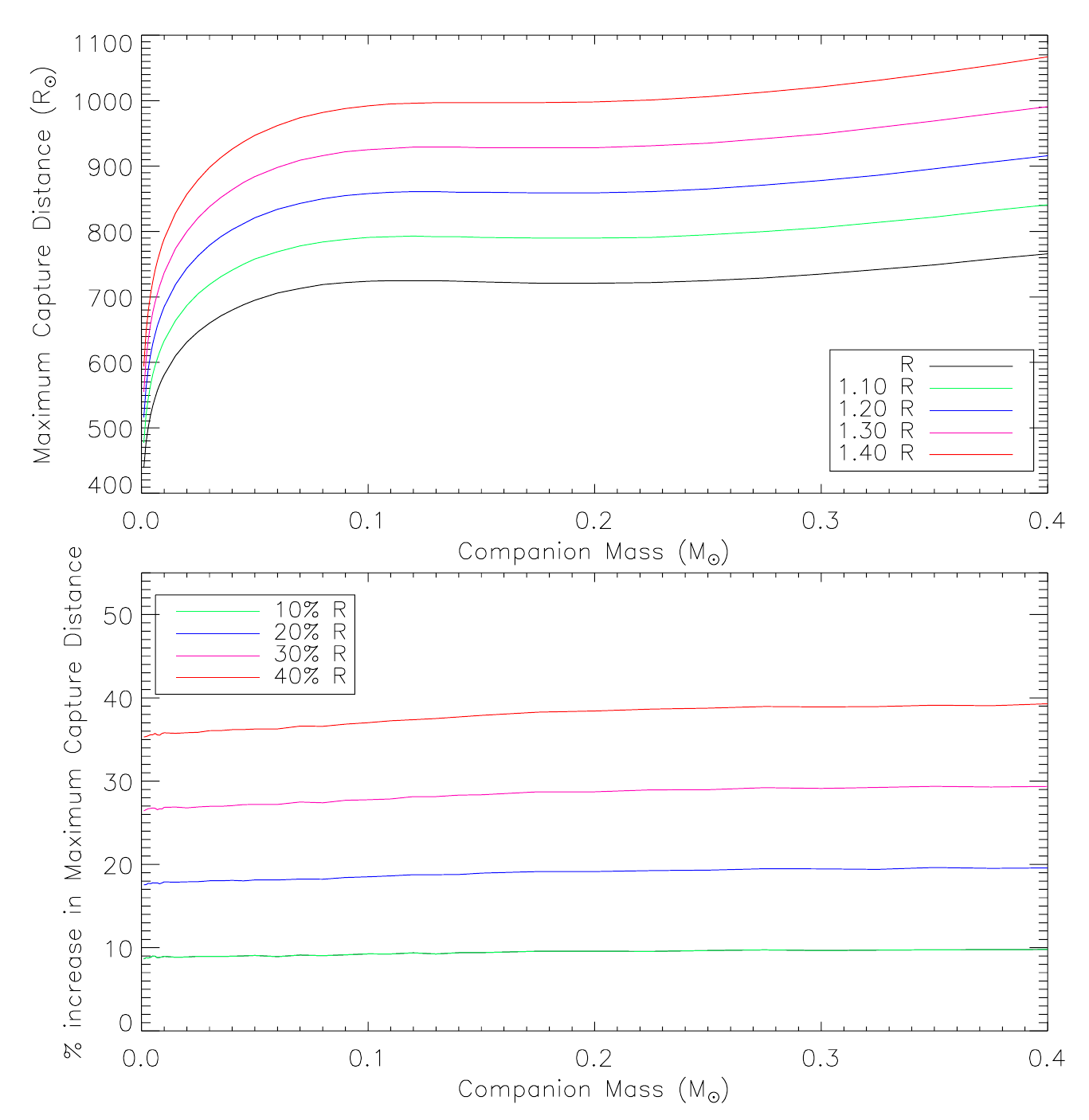}
\caption{{ The AGB MCD (top panel) and percentage difference (bottom panel) for a 1.5~\msun\ stars that has its radius artificially increased between 10 and 40 per cent.}}\label{fig:R_Uncert}
\end{figure} 

\subsection{The Maximum Capture Distance During the RGB or AGB} 
\label{section:Acap_RGB}

The top panel of Fig. \ref{fig:Acap_RGB} shows the MCD as a function of companion and primary ZAMS mass. The primary mass is between 0.8 and 4.0 M$_\odot$. The  companion mass range is between 0.001 ($\sim$1 M${\rm{_J}}$) and 0.78 M$_\odot$. In the bottom part of Fig. \ref{fig:Acap_RGB} we display the same quantity but as a function of companion-to-primary mass ratio and primary mass, where for each primary a range of simulations was carried out with companion masses up to a mass ratio of just below unity. The most massive binary in our calculation has $M$ = 4~M$_\odot$ and M$_2$ = 3.9~M$_\odot$ for Z = 0.01.
  
Fig. \ref{fig:Acap_RGB} shows two distinct areas. The first region is a rectangle between primary masses 0.8 and 1.8 M$_\odot$. In this mass range the peak RGB radius is between $\sim$120 and $\sim$150~R$_\odot$ (for the 0.8 and 1.0~\msun, primaries, respectively; see Fig. \ref{fig:MultiRadcom} for Z=0.01). As a result of such high peak RGB radius, the MCD is also high. Within this primary mass range, the MCD initially increases as a function of mass ratio, up to $q=0.05$ to values of $\sim$470 R$_\odot$. For mass ratios larger than $\sim$0.05 we see a sudden decrease in the MCD to $\sim$300~\rsun, that can be attributed to the tidal spin up of the primary. At  q\,$\sim0.4$, the MCD starts to increase again gradually, particularly for the lower mass primaries, due to a larger value of $q$ in Eq.~\ref{eq:newaa1}.  In conclusion, primaries with ZAMS mass $\lesseq$1.8~\msun\ capture companions that are typically closer than $\sim$2-3 times their maximum radius, with the lowest mass companions being captured out to almost 4 times the maximum radius.

Between ZAMS masses of 1.8 and 2.0 M$_\odot$ we see in Fig. \ref{fig:MultiRadcom} a sudden drop in the peak RGB radius to $\sim$40~R$_\odot$, leading to a drop in MCD in Fig. \ref{fig:Acap_RGB}. Between ZAMS masses of  2.0 and 4.0~M$_\odot$ the peak RGB radius sits between 40 and 60~\rsun, resulting in a MCD between $\sim$40 and 200~\rsun. We also witness the initial increase, peak and decrease behaviour as a function of mass ratio also seen at lower primary masses. In conclusion, primaries with ZAMS mass $>1.8$~\msun\ capture companions that are typically out to $\sim$1-2 times their maximum radius, with only the lowest mass companions being captured out to almost 4 times the maximum radius. 

Fig. \ref{fig:Acap_AGB} shows the MCD during the AGB phase as a function of primary ZAMS mass and secondary mass (upper panel) or mass ratio (lower panel).  At a metallicity of Z=0.01, a primary star of 0.8 M$_\odot$ forms a helium white dwarf and does not ascend the AGB. So the AGB MCD for these systems is assumed to be zero. Maximum AGB radii increase with ZMAS and are between 220 and 700~\rsun, for masses of 1 and 4~\msun, respectively. Companions are captured out to a range of distances between 350 and 2500~\rsun. For each ZAMS mass the capture distance increases with companion mass, although such increase is mitigated by spin-orbit coupling at intermediate companion masses. 

Overall AGB stars in the 1-4~\msun\ mass range capture companions that are between 1 and 4 times the maximum AGB radius. The lowest mass companions, such as planets, only get captured out to 1-1.5 times the maximum AGB radius. The most massive primaries are unable to tidally capture planetary companions, but they capture massive companions ($q\sim1$) out to almost 4 times the maximum AGB radius. The 2-\msun\ primaries capture companions the farthest relative to their radii, but, in absolute terms, it is the 4-\msun\ primaries that capture companions the farthest. The implication of this fact will be discussed in Section~\ref{sec:summary}. 
\begin{figure}
\includegraphics[width=90mm]{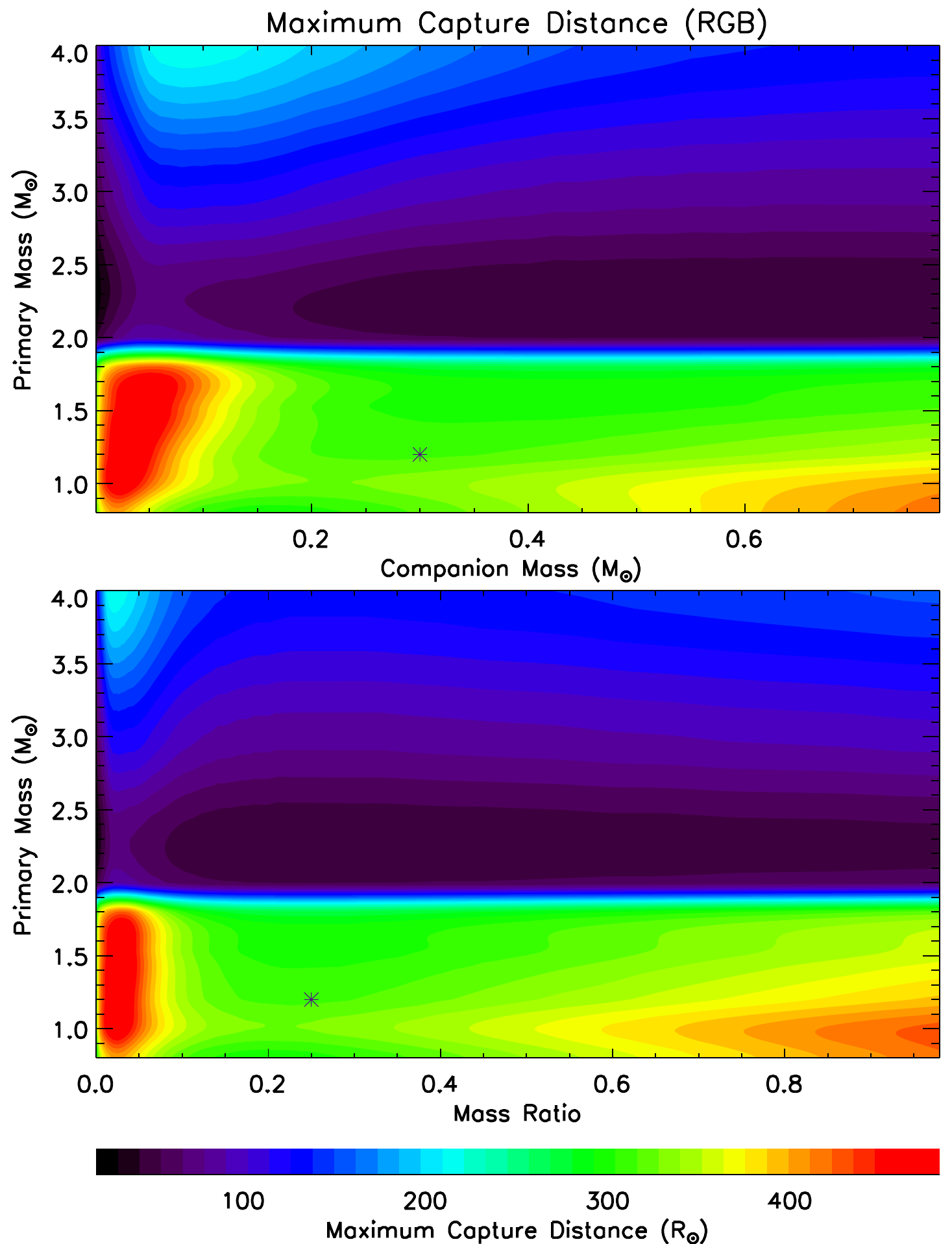}
\caption{The MCD as a function of primary and companion mass (top panel), or mass ratio (bottom panel) for the RGB. Stellar models are for $Z=0.01$. The star symbol denotes a binary system with a 1.2-M$_\odot$ primary and a 0.3-M$_\odot$ companion.}\label{fig:Acap_RGB}
\end{figure} 
 \begin{figure}
\includegraphics[width=90mm]{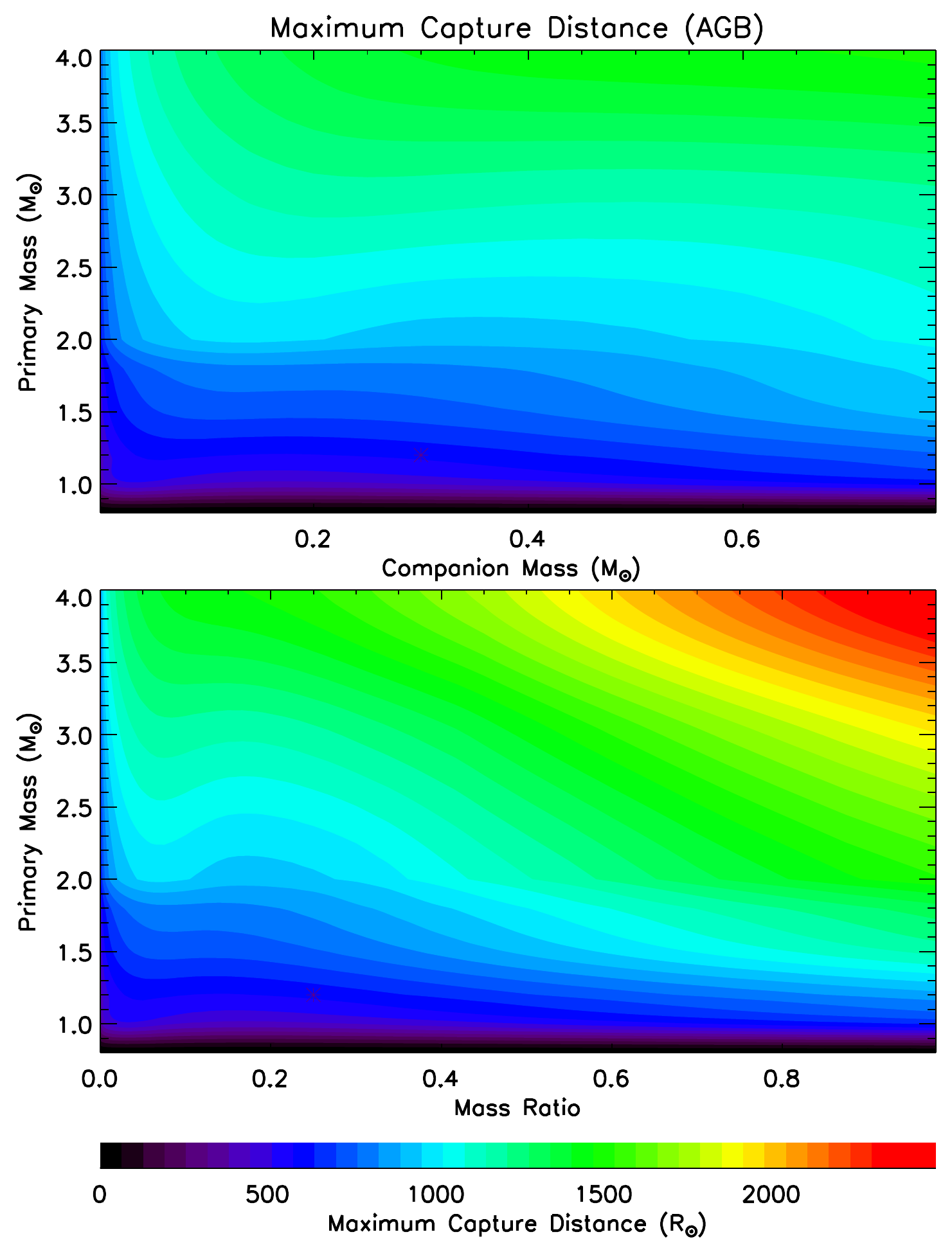}
\caption{The MCD as a function of primary mass and companion mass (upper panel) or mass ratio (lower panel) during the AGB. Stellar models are for $Z=0.01$. The star symbol denotes a binary system with a 1.2-M$_\odot$ primary and a 0.3-M$_\odot$ companion.}\label{fig:Acap_AGB}
\end{figure}

\subsection{Which Binaries Interact on the AGB}
\label{section:keyhole}

In Figs. \ref{fig:Acap_RGB} and \ref{fig:Acap_AGB} we plot an asterisk at $M = 1.2$~\msun\ and $M_2$ = 0.3~\msun. These are typical values for the ZAMS mass of a PN central star and the mass of its companion. These values are selected in the following way. From a population synthesis calculation, the PN progenitor mass distribution peaks steeply at 1.2~\msun\ \citep[][their figure 10]{MoeDeMarco2006}, despite the initial mass function peaking at lower masses \citep[e.g.,][]{Chabrier2003big}, because stars with a mass lower than $\sim$0.9~\msun\ tend not to make visible PN\footnote{ This is known as the ``lazy PN" paradigm, discussed, e.g., by \citet{Jacoby1997} and also explained in \citet{Moe2006}, their figure 7, section 3.7 and references therein.}. As for the companion mass, if the close companions to the central stars of PN had the same spectral type distribution as the companions to white dwarfs, then the mean spectral type of the companions to central stars of PN would be $\sim$M3V \citep[this is the most represented spectral type in the histogram of WD companion spectral types of][]{Farihi2005}, corresponding to masses of 0.33~M$_\odot$ \citep{Raghavan2010,DeMarco2013}. 

Hence the asterisk in Fig.~\ref{fig:Acap_AGB} informs us that for a representative PN central star binary companions orbiting farther than approximately 880~R$_\odot$ will not be captured into an AGB interaction. On the other hand, the asterisk in Fig.~\ref{fig:Acap_RGB} informs us that companions closer than $\sim$320~\rsun\ will interact during the RGB. Those systems that interact via a CE on the RGB and survive as binaries are unlikely to ascend the AGB. This is due either to a low envelope mass that would prevent an AGB ascent \citep{Dorman1993} or to the fact that if the post-RGB primary attempts to expand on its AGB ascent, it will suffer a CE with its very close companion, preventing further expansion. In conclusion, in the mass range 1-4~\msun\ and for $Z=0.01$, only binaries with orbital separations between $\sim$320 and $\sim$880~\rsun\ will enter a CE on the AGB. 

Instead than simply adopting MCD values corresponding to one representative primary-companion mass combination we actually  convolved the columns of Figs.~\ref{fig:Acap_RGB} and \ref{fig:Acap_AGB} with the function representing the progenitor mass distribution for PN \citep[][their figure 10]{MoeDeMarco2006} and the rows with the companion mass distribution from \citet{Raghavan2010}, which is flat (all mass ratios are equally represented). From this exercise we obtain the same MCDs of 320~\rsun\ for the RGB and a revised value of 630~\rsun\ instead of 880~\rsun\ for the AGB.  }

\begin{figure}
\includegraphics[width=80mm]{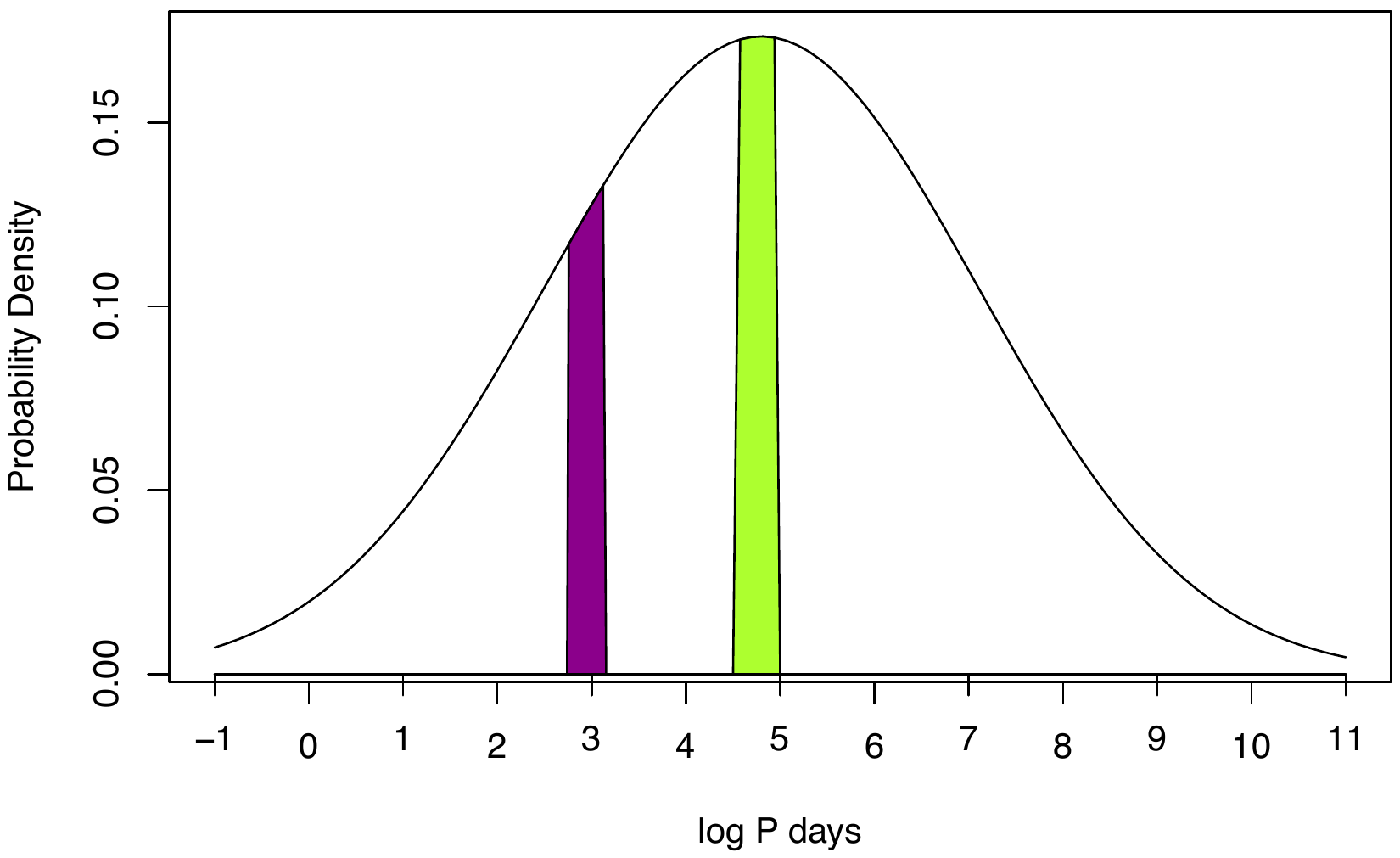}
\caption{The normalised period distribution of F and G main sequence stars from \citet{Raghavan2010}. The purple area designates those binaries (5 per cent of all binaries, or 2.5 pr cent of all systems) that will interact on the AGB via a Roche-lobe-filling interaction as calculated in this work. The green area demonstrates that even { increasing the strength of the tide so as to capture companions farther out, where there are more companions, the predicted fraction of binaries that will interact on the AGB only increases from $\sim$5 to $\sim$9 per cent (see text in Sec.~\ref{ssec:fraction}).}}
\label{fig:percentages} 
\end{figure}

Looking at the period distribution of main sequence binaries of  \cite[][Fig.~\ref{fig:percentages}]{Raghavan2010}, we see that 5 per cent of main sequence binaries have orbital separations between 320 and 630~\rsun. For a 50 per cent binary fraction for the PN progenitor population\footnote{Note that this fraction is reasonably accurate, with the value 50$\pm$4 per cent determined by Raghavan et al. 2010 for the F6V-G2V primaries that are most relevant here where WD as well as brown dwarf companions were observed or accounted for in their completeness study} \citep[][]{Raghavan2010}, we expect {$\sim$2.5 per cent} of PN would have a CE origin. 

For completeness we should also discuss the actual shape of the mass radio distribution, $q$, which we have adopted to be flat in the calculation above. For the G type stars, the distribution is flat for the wide binaries, but favours $q$ values close to unity for the close binaries of interest here, with the exponent of the distribution close to $\gamma = 1.2$ ($f(q)=q^{\gamma}$; see figure~3 of \citet{Duchene2013}). Adopting such a distribution may, however, not be entirely appropriate. The spectral type of the known main sequence companions to post-CE central star primaries is, with very few exceptions, MV. This is due to the discovery method of post-CE central stars, which does not readily detect bright, hot companions. Hence, adopting a mass ratio distribution that favours higher mass companions may in the end misrepresent the population of known post-CE central stars. In the end, however, this argument is mute, because the predicted percentage post-CE central stars would only increase by less than 1 percentage points even if {\it all} companions were as massive as the Sun.   

We must remember that we have carried out this calculation at $Z=0.01$. At twice that metallicity (the Solar metallicity is $Z_\odot$=0.014) the RGB and AGB stellar radii are $\sim$10 per cent larger and the MCDs would increase accordingly (Section~\ref{section:R_Uncert}) to $\sim$350 to $\sim$690~\rsun, which leaves effectively unaltered the fraction to 2.5 per cent. The predicted percentage of post-CE central stars of PN is, at $\sim$2-3 per cent, much smaller than the known post-CE binary fraction for central stars of PN of 15-20 per cent \citep{Bond2000, Miszalski2009}, which is itself a lower limit. We discuss the interpretation of this discrepancy in Section~\ref{ssec:fraction}.

\section{Comparison with other tidal prescriptions} 
\label{sec:Hurley2002_com}

Below we compare our results the work of \citet{Mustill2012} and \citet{Nordhaus2013} calculating the tidal interaction between intermediate mass stars and their planetary companions. Afterwards we compare our work with the tidal models of \citet{Hurley2002}, commonly used in the case of stellar mass companions. 

\citet{Villaver2009} and \cite{Mustill2012} studied the capture distance for the range of primary ZAMS masses 1-5~\msun\ orbited by Jovian and  Neptunian-mass companions. The first difference between our and their models is the use of different stellar evolutionary calculations (\citet{Villaver2009} used STAREVOL \citep{Siess2006} for their RGB calculations and \citet{Mustill2012} based their stellar parameters on \citet{Vassiliadis1994} and \citet{Karakas2002} for their AGB calculations). 
On the RGB their maximum radii are 60 per cent larger for lower mass stars but very comparable for more massive stars. On the AGB their maximum radii are approximately 50 per cent larger than ours. Their MCDs on the RGB and AGB are larger than ours by a factor similar to that for our respective maximum stellar radii.
Also, the dip in maximum RGB radius for their calculations happens at slightly larger masses than for ours. Villaver et al. (2014) have shown how changing slightly the RGB mass-loss prescription in the stellar evolution calculations results in a highly non-linear process that modifies the stellar  radius and thus the capture distance. The differences between the MCDs presented in the models of \citet{Mustill2012} and those presented here can be attributed to the different stellar models used and the fact that the stellar structure is accounted for in their calculation of the tidal forces. For a full description of how changes in the parameters entering the calculation of the tidal forces affect the capture distance see \citet{Mustill2012}.


\cite{Nordhaus2013}'s secular tidal equations are essentially similar to our own, but for the fact that they have  
eliminated $k_2$ and $f$ on the grounds that their values are close to unity. The apsidal constant, $k_2$ is the same as $\lambda_2$ and $\lambda_2\,=\,f/21$, where $f$ is of order unity; hence $k_2$\footnote{\cite{Nordhaus2013} have called $k_2$ the tidal love number, which is twice the apsidal constant \citep{Sterne1939}.} cannot be unity.  Values of $k_2$ for stars with extended convective envelope have been measured to be between 0.1 and 0.003 \citep{Torres2010} 
and are seen to vary during the evolution of a star between 0.001 and 0.063 (see figure 2 of \citealt{Claret2004}). Our equations, which are similar to those adopted by \cite{Hurley2002}, therefore result in a factor in Equation~\ref{eq:newaa1} that is $\sim$20 times smaller than in \cite{Nordhaus2013}.

\begin{figure}
\includegraphics[width=92mm]{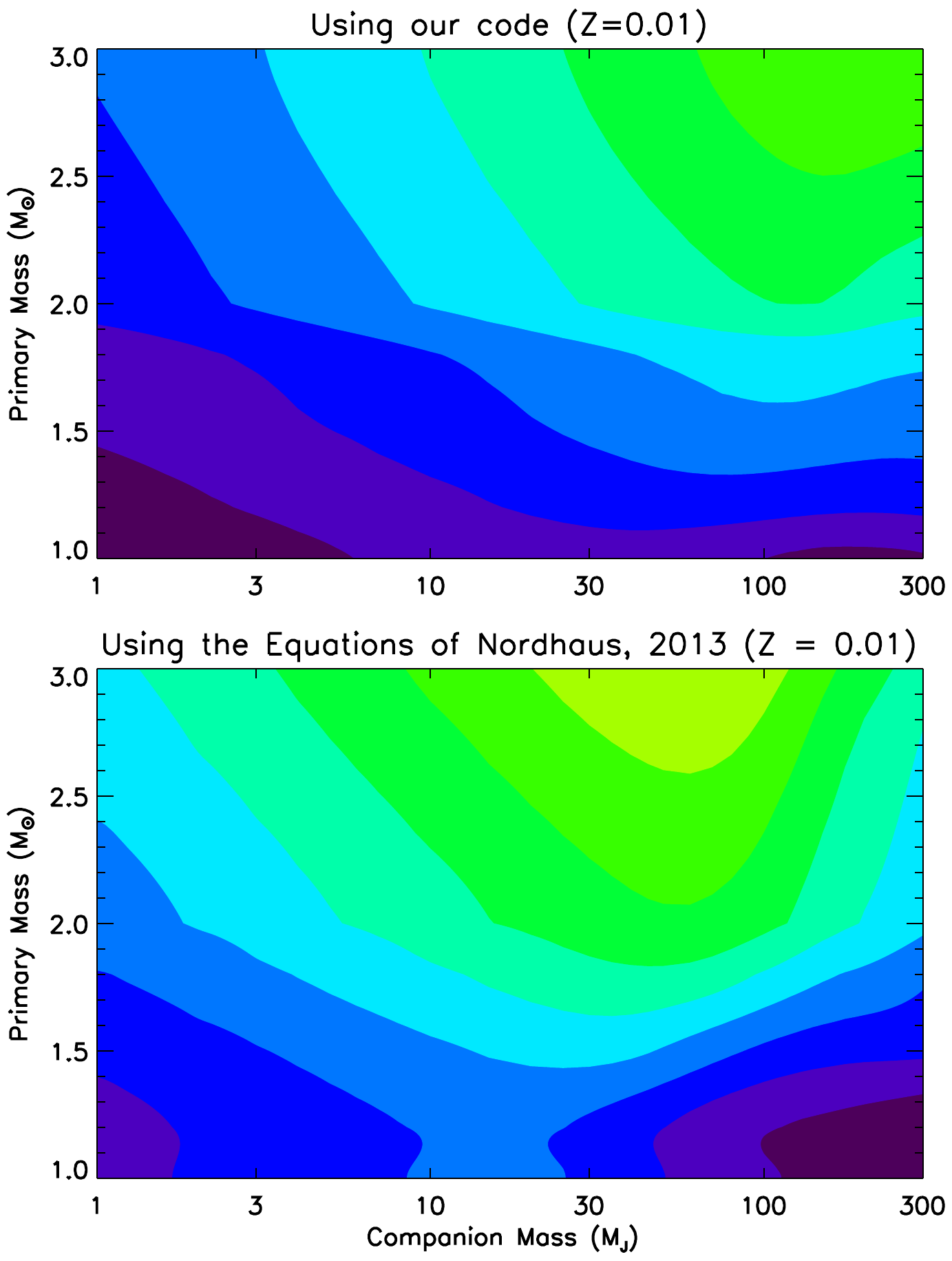}
\includegraphics[width=90mm] {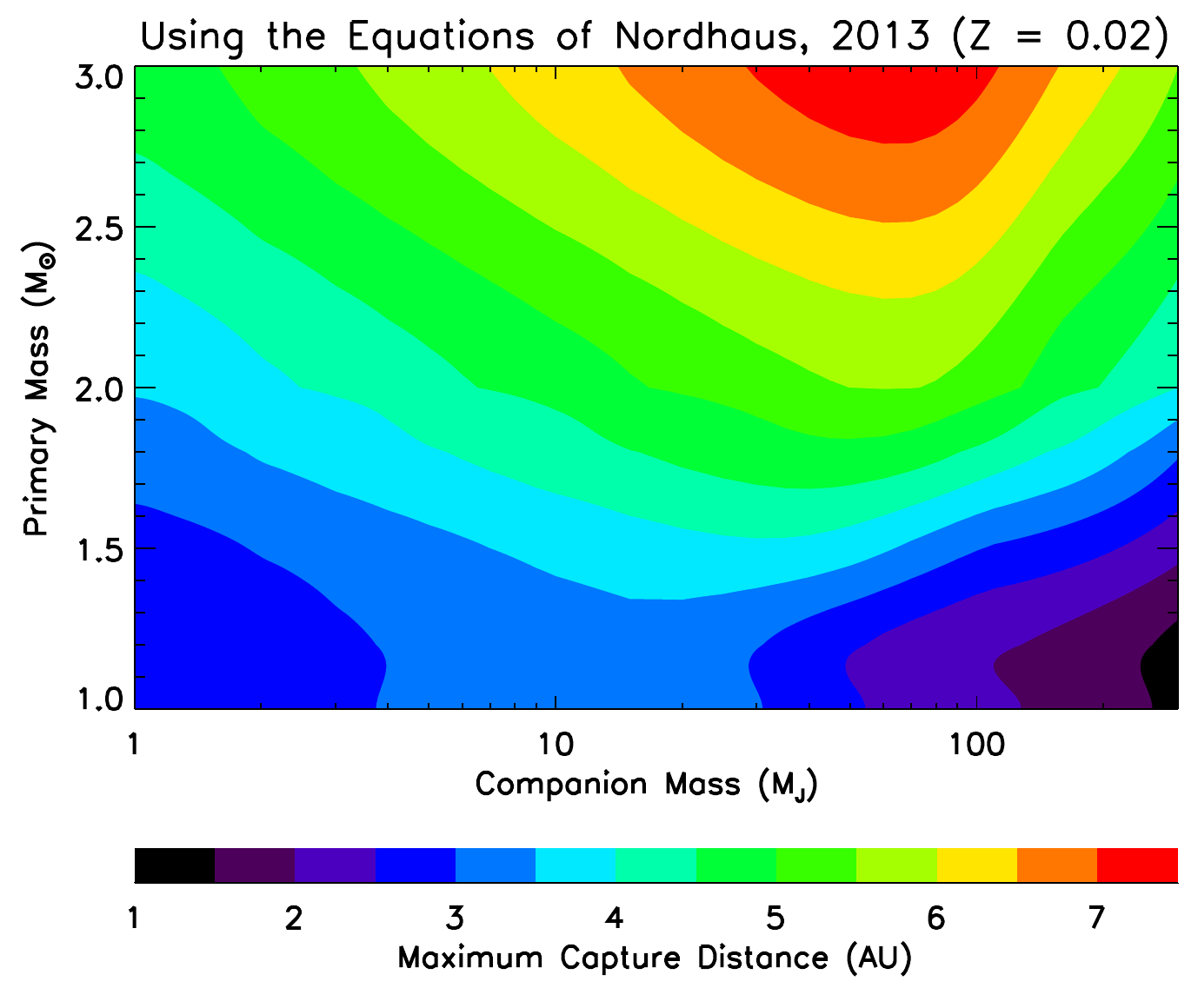}
\caption{The MCD as a function of the primary mass and the mass ratio for primary stars with ZAMS metallicity of Z= 0.01 using our prescription (top panel). This is the same plot as in Fig. \ref{fig:Acap_AGB} but with different $x$ and $y$ ranges. Middle panel: the same contour plot, but this time produced using the tidal equations of \protect\cite{Nordhaus2013}. Bottom panel: the same plot as in the middle panel, but with a metallicity Z=0.02.}
\label{fig:Contour_Z01} 
\end{figure}

This said, the MCDs obtained by \citet[][their figure 3]{Nordhaus2013} are only between 1.4 and 1.9 times larger than ours (Fig. \ref{fig:Contour_Z01}; top panel), with the least difference happening at lower primary, but higher companion masses. Assuming their metallicity to be approximately Solar, part of this difference could be ascribed to our using a lower metallicity value for our stellar model. To disentangle the effect of larger stellar radii from stronger tides, we calculated MCDs using the equations of \citet{Nordhaus2013} with $Z=0.01$ (Fig.~\ref{fig:Contour_Z01}, middle panel) or $Z=0.02$ (Fig.~\ref{fig:Contour_Z01}, lower panel).  Comparing the uppermost with the two lower panels shows that the difference in MCDs is reduced to between 1.2 and 1.6 (middle panel) and to between 1.02 and 1.5 (lower panel). From this we deduce that the larger factor in Equation 1 used by \citet{Nordhaus2013} and the larger metallicity account approximately in equal measures for their larger capture distances. Any remaining differences (mostly at larger primary masses) could be accounted for by stellar structure model differences, including the fact that during the upper AGB we terminated the models early due to convergence issues, likely resulting in slightly lower maximum radii  in that phase. 

Our work can also be compared with the star-planet tidal calculations of \citet{Carlberg2009}, who used an earlier version of the Padova stellar evolution models \citep{Girardi2000} up to the RGB phase of evolution. Their RGB MCDs for a 1-\mjup\ mass companion are 20 per cent smaller than ours for 1~\msun\ stars and are almost identical to ours  for $\sim$1.5~\msun\ stars. The analytical calculations of \citet{Adams2013} are said by them to be comparable to those calculations carried out numerically, among which they list \citet{Nordhaus2013} and \citet{Mustill2012}.

\begin{figure}
\includegraphics[width=88mm]{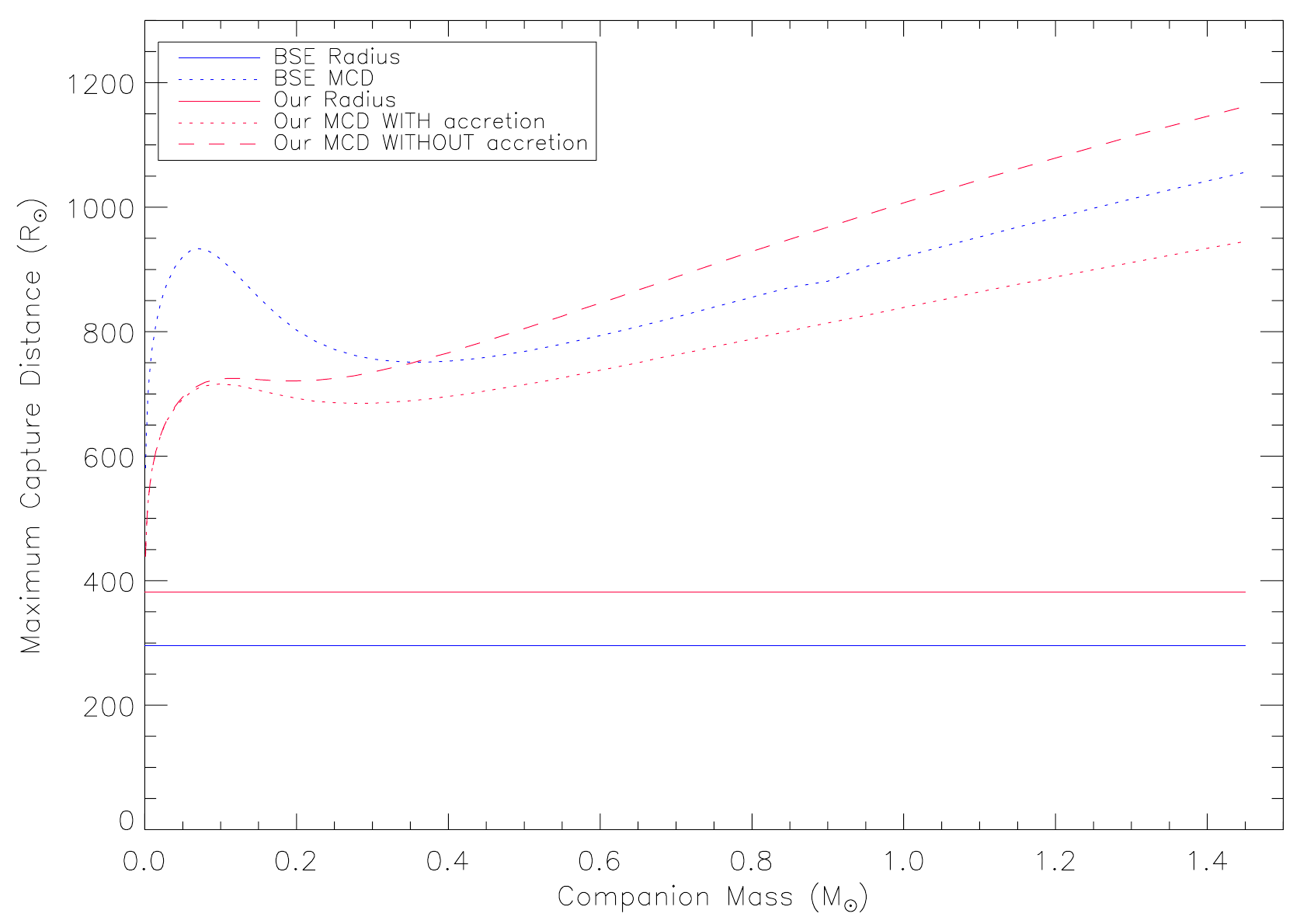}
\caption{The AGB MCD as a function of companion mass for a primary of 1.5 M$_\odot$. The MCD is calculated for the AGB stage using our code with (pink dashed line) and without (red dotted line - lower dotted line) accretion onto the companion and the BSE code (with accretion onto the companion; blue dotted line - upper dotted line). The red (upper) and blue (lower) solid horizzontal lines are the peak AGB radii using MESA and BSE, respectively. \label{fig:M1M2_hurley}}
 \end{figure}

Finally we compare our results with the work of \cite{Hurley2002} by using the publicly-available BSE code\footnote{astronomy.swin.edu.au\/\~jhurley\/bsedload.html}, but setting the companion-induced mass loss to zero.  Fig. \ref{fig:M1M2_hurley} compares the MCD obtained using BSE and our code with  a 1.5 M$_\odot$ primary with Z=0.01 and a range of companion masses. 
If no accretion onto the secondary is included, the AGB MCD obtained using our code is 
approximately 1.1-1.3 times smaller than using BSE, even though the stellar evolution model calculated using MESA results in a peak AGB radius that is 1.2 times {\it larger} than that obtained using the fitting formulae in BSE.
Including accretion onto the secondary, our model's AGB MCD for companions more massive than $\sim$0.1~\msun\ is larger. Our MCD remains below the BSE value up to companions of 0.4~\msun, but exceeds it for companions with a larger mass, becoming 1.1 times larger than the BSE value for companions with 1.4~\msun. For the lower mass companions the difference can be attributed to the higher mass-loss rate in the MESA  AGB models used by our tidal code. The BSE code uses a \cite{Reimers1975} mass-loss prescription, adequate for the RGB case,  but predicting too low a mass-loss rate on the AGB. Instead MESA uses \cite{Bloecker1995}'s AGB mass loss prescription, which accounts for the enhanced mass loss during this stage of evolution. Mass loss acts to increase the separation between the primary and companion (see Equation \ref{eq:newaa1}, Section \ref{sec:Zahnnew}), resulting in farther companions avoiding capture. For the more massive companions mass-loss has a smaller effect, and the larger MCD predicted by our work can be attributed to the thermal pulse radius meeting the Roche radius. Note in fact that maximum inter-pulse radius of our 1.5-\msun\ model is very similar to that of the fitting formulae.

{ Fig.~\ref{fig:M1M2_hurley} can also be used to make the following consideration. The largest difference between including and excluding accretion onto the companion is $\sim$20 per cent at larger companion masses. \citet{HuarteEspinosa2013} reported that for smaller orbital separations, accretion rates tend to be below those expected using the Bondi-Hoyle approximation (they simulated orbital separation of 10, 15 and 20~au, larger than any of the values we consider). If we have overestimated accretion onto the companion, then we would have overestimated the capture distances during the AGB (the RGB values would be minimally affected, because of the much lower mass-loss rates of those stars). As a result we would also have overestimated the fraction of central stars that have gone through an interaction.}   

We finally point out that the differences we have encountered between our work and both planetary companion studies as well as BSE would be insufficient to alter our conclusion of a very small predicted percentage of post-CE PN.

\section{Summary and Discussion}
\label{sec:summary}

We have integrated the tidal equations of \citet{Zahn1977,Zahn1989} assuming zero eccentricity in order to determine how far out intermediate mass stars (ZAMS mass between 0.8 an 4.0~\msun) capture their planetary and stellar companions into strong interactions that would lead to mass transfer or CE interactions. The aims of our study was to understand how often binary interactions take place during the upper AGB, something that will affect the PN shape.

Our tidal integration combined with the MESA models lead to slightly smaller capture distances than past work for planetary mass companions, with most of the differences ascribed to the use of stellar models with different radial evolutions. Such differences are not large when we consider the overall uncertainties concerning tides. Here we have compared our code also to that of \citet{Hurley2002} who included spin-orbit interaction. Our model predicts capture distances that are approximately 30 per cent lower for lower mass companions and 10 per cent larger for more massive companions. These differences are ascribed to the different mass-loss prescription of  fitting formulae, for the lower mass companions, and the lack of thermal pulses for the higher mass companions.

The MCDs predictions of Figs.~\ref{fig:Acap_RGB} and \ref{fig:Acap_AGB} are suitable for integration into a population synthesis model of intermediate mass stars as done for example by \citet{Moe2012}.

\subsection{The Evolution of the Stellar Radii}
\label{ssec:stellarradii}

We have first presented a comparison of the maximum RGB and AGB radii reached by stellar models with ZAMS mass smaller than 4.0~\msun, between a calculation carried out with the code MESA and calculations carried out using the Padova models of  \cite{Bertelli2008, Bertelli2009}, as well as by \cite{Hurley2000} using fitting formulae. Differences between these approaches propagate into tidal captures results. 

We first remark on the strong dependence of the RGB maximum stellar radius on metallicity for ZAMS masses smaller than $\sim$2~\msun, where the highest metallicities experience the largest radii (Fig.~\ref{fig:MultiRadcom}). This would dictate a relatively larger fraction of RGB binary interactions in higher metallicity environments, such as our Galaxy compared to, for example, the Small Magellanic Cloud galaxy. Relatively more RGB interactions translates to relatively fewer AGB interactions. Hence at higher metallicities, the lower mass binary population contributes to making PN {\it less} than the higher mass population. Hence PN from binary interactions would have a higher mean mass in the Galaxy than in the Magellanic Clouds. This may already have been born out by observations \citep{Moe2006}, although the uncertainties in deriving central star masses may well confuse this test.

Second, we see that  MESA models and the  fitting formulae predict smaller RGB maximum radii for ZAMS masses smaller than $\sim$2~\msun\ than the Padova models of \citet{Bertelli2008}. Maximum RGB radii for higher mass stars and maximum AGB radii do not vary systematically, nor significantly as a function of metallicity or even for different models and any scatter is due to a combination of other issues rather than systematic trends. Therefore the only systematic and substantial repercussion of implementing different stellar models in population synthesis studies would be for the predicted rates of RGB interactions for lower mass primaries. Since lower mass, PN-making stars are more numerous than their more massive counterparts, using the Padova models of \citet{Bertelli2008,Bertelli2009} instead of MESA should lead to fewer PN from binary interactions overall and a higher overall mass for the central stars from binary interactions.

One of the motivations of this paper was to determine whether the sudden and large increase in the stellar radius during an AGB thermal pulse translates into a powerful tidal force. Had this been so, it would result in capturing companions out to larger distances compared to the MCD calculated using of a stellar model with no thermal pulses. This is an important question, considering that many stellar population synthesis codes use the \citet{Hurley2000} stellar structure fitting formulae, which do not include thermal pulses. The answer is that thermal pulses have too short a duration to result in increased capture distances. As a result, the MCD is more influenced by the maximum {\it inter-pulse} radius than by the pulse radius. 

However, the occurrence of a thermal pulse can result in a {\it direct} capture, if the pulse radius fills the Roche lobe of the star resulting in mass-transfer. In such an eventuality, not including the pulse radius in a tidal calculation would result in no capture for the same parameters. This is well depicted in Fig.~\ref{fig:Avst2}, where the farthest companion that is captured would have escaped capture had it not been for the last thermal pulse. Interestingly, if the tidal interaction is very weak (low mass companions), all captures happen during thermal pulses, as seen in the work of \citet{Mustill2012} where Earth-mass companions are captured on the AGB exclusively during pulses. The more massive companions have instead a higher chance of being captured in the inter-pulse phase, as the strong tide shortens the orbital separation leading to a Roche lobe overflow before the next pulse. 

AGB stars suffer pulsations during the end of their evolution (the Mira phase) that increase the radii by as much as 25 per cent with periods of the order of one hundred days \citep[e.g.,][]{Ireland2011}. These oscillations are not included in the stellar evolution models used here. If they were, the time-averaged radius during these phases would be larger and the spin frequency of the primary, proportionally smaller. These oscillations are short lived and would not alter, per se the tidal interaction. However, on average we could expect a tidal action proportionally stronger, because of the larger mean radius. The lower average spin frequency of the primary, would contribute to lower the synchronisation factor (see Fig.~\ref{fig:Synchro_com}), something that would not per se alter the tidal strength particularly, but could play a role for those cases where the synchronisation factor is close to unity at those times when the tide is strong, for example in the inter-pulse phase (this is the case for our 0.15~\msun\ companion in the bottom panel of Fig.~\ref{fig:Synchro_com}). Such an effect would be more important for more massive companions than lower mass ones for which the synchronisation factors are in any case well below unity.

\subsection{The fraction of PN with post-common envelope binary central stars}
\label{ssec:fraction}

In Section~\ref{section:keyhole} we have concluded that main sequence star binaries in the mass range 1-8~\msun\ result in a post-CE PN if their companions are in the range $\sim$320 -- 630~\rsun. Using the 50 per cent binary fraction \citet{Raghavan2010}, we have predicted that only 2.5 per cent of the population should have gone through a Roche lobe filling interaction on the AGB.  If all such interactions lead to a CE and all those CE lead to a post-CE binary, then 2.5 per cent of all PN should have such a post-CE binary in their centres. Some of the Roche-lobe filling interactions will not lead to a CE interaction and some CE interactions will lead to a merger instead of a binary. Hence $< 2.5$ per cent of all central stars of PN should be post-CE binaries today.  Contrary to this prediction, observations indicate a fraction of post-CE binaries of {\it at least} 15-20 per cent. 

Most numbers used in our calculation (e.g., the period distribution, or the binary fraction) have uncertainties that are not large enough to alter the prediction significantly. By far the largest uncertainties in our calculation reside in the maximum AGB radii and on the treatment of tides. This was, after all, the reason for the current paper: a tidal calculation with a better understanding of spin-orbit coupling using AGB stellar evolutionary models that include the thermal pulses. We may therefore wonder whether these uncertainties could explain away the discrepancy between our prediction and the observed fraction of post-CE central star binaries. 

Looking at the period distribution in Fig.~\ref{fig:percentages} we see that by making the tides an unrealistic 10-15 times stronger we would still only achieve a fraction of post-CE PN of 4.5 per cent of all stars (green area in Fig.~\ref{fig:percentages}), still below the required observed fraction.

We can also work in reverse, and ask what MCDs are needed to increase the predicted percentage to match the observed 15-20 per cent of all stars, (30-40 per cent of all binaries). The answer is that we would have to increase the purple area by keeping its left boundary similar (similar tidal strength and similar maximum RGB radii),  but shifting the right boundary to larger values of $log(P)$ on the assumption that the maximum stellar radii during the AGB are much larger than we have currently calculated. In order to match the observed fraction, the maximum AGB radii would have to be larger by a factor of $\sim$10, which is unlikely even given the large uncertainties on AGB radii. There seems to be no reasonable way to increase the predicted fraction of captured companions even to 30 per cent, the lowest bound of the observational lower limit. We conclude that the uncertainty concerning the treatment of tide nor on the prediction of AGB radii would not easily reconcile the observations with the prediction.

Past population syntheses predicted rates that are larger than the current estimate, but also too low compared to the observations. \citet{Han1995}, who did not include tides, predicted a percentage of PN with post-CE central stars below 9 per cent of all binaries, or 4.5 per cent of all stars, using a 50 per cent binary fraction. Their not having included tides may have resulted in a larger range of binary separations that lead to post-CE PN (see Sec.~\ref{sec:Planet}). On the other hand, \citet{Nie2012}, who used a different tidal prescription and a novel and completely independent method, calibrated using the fraction of Large Magellanic Cloud RGB stars that has a close companion, predicted tyhat 7--9 per cent of all PN contain a post-CE binary, for their preferred models. However, their preferred model also predicts a main sequence binary fraction of 81--97 per cent, higher than the observational determination of Raghavan et al. (2010) by a factor of two. Choosing one of their other proposed models with a predicted binary fraction more in line with observation, the fraction of post-CE central stars is 5-6 per cent. Both these studies predict post-CE central star populations that are a factor of two larger than ours, but still 3-4 times smaller than the observational lower limit.

\subsubsection{Explaining the discrepancy between predicted and observed fraction of post-CE PN}
\label{sssec:explanation}

 Here we discuss possible explanations of why the predicted fraction of post-CE PN is lower than the observational lower limit. 

A first way to reconcile the two numbers is by assuming that the visibility time for post-CE primaries is more than 6-8 times longer than the visibility time of single central stars. There are two ways to increase the visibility time of a PN: a slower PN expansion speed and a lower mass central star combined with a faster central star transition time (the time taken by the post-AGB star to reach a temperature of approximately 25\,000~K sufficient to ionise the PN). There is no real reason (and no evidence) why the PN around post-CE binaries should expand more slowly. 

On the other hand, post-CE central star could in principle have a longer visibility time in virtue of a systematically lower mass and short transition time. The CE presumably interrupts the AGB evolution and concomitant core growth, producing a lower mass central star. \citet{MillerBertolami2016} calculated that a 1.5~\msun\ star has a core mass of 0.53~\msun\ at the first thermal pulse, but 0.58~\msun\ at the end of single-star AGB  evolution. This could be the mass difference between a post-CE central star and a single one, derived from the same main sequence mass. The time between AGB departure and maximum central star effective temperature is 25\,000~yr for the 0.53~\msun\ model of \citet[][called the crossing time by them]{MillerBertolami2016} at Z=0.02, while for a 0.58~\msun\ model it is 4500~yr. The  transition times of a post-CE central star is likely nil: the CE in-spiral reduces the cool AGB giant to a small star with the same luminosity and hence a high effective temperature, within a dynamical time ($\sim 1$~yr). The transition time of a 0.58~\msun\ star is 3400~yr. Assuming that the visibility time is the crossing time minus the transition time, it is likely that the visibility time of the lower mass, presumably post-CE central star is 20 times longer. This could in principle explain an over abundance of post-CE central star. This said, there is no real evidence that post-CE central stars have systematically lower masses from either binary modelling \citep{DeMarco2008,Hillwig2010,Hillwig2015} or other types of measurement \citep[e.g.,][]{Stasinska1997}, though the relatively few measurements are fraught with a large range of diverse uncertainties. 

An alternative explanation is that the observed overabundance of post-CE binaries is due to interacting stars making PN in preference to non-interacting stars, with the latter group making under-luminous, hard to detect PN. This would  inflate the interacting binary fraction in the PN population.  This point has already been argued by others on different lines of evidence \citep{Soker2005,Moe2006,Jacoby2010,Moe2012}. 

This hypothesis makes testable predictions.  Stars from slightly more massive progenitors ($>$2.5~\msun) will capture their companions on the AGB if their initial orbital separation is between 30 and 1000~\rsun. Twenty-two per cent of the main sequence binary population has orbital separations in this range. Stars with a ZAMS mass of 2.5~\msun\ are A0V stars, which have a binary fraction of $\sim$70 per cent \citep{Raghavan2010}, larger than the 50 per cent known for the F and G stars. Hence, the fraction of central stars in post-interaction binaries for this group would be $\sim$15 per cent. This means that the more massive population has a much larger chance to interact with a companion, leading to the prediction that the PN population from binary interactions has more massive central stars on average than, say, the post-AGB white dwarf population \citep{Liebert2005}, which doubtless derives from {\it single and binary} $\sim$1-8~\msun\ stars. 

Qualitatively the above prediction may already have been born out: bipolar PN must derive from relatively more massive stars because of their low scale height \citep{Corradi1995} and because a large fraction of them has a high N/O ratio (indicative of ZAMS mass $\gtrsim$4.0~\msun; \citet{Kingsburgh1994}).  At the same time, bipolar PN are increasingly being associated with close binary central stars \citep{Soker1997,DeMarco2009,Miszalski2009b}. This points to the population of post-interaction binary central stars having more massive central stars, as predicted. A quantitative prediction needs a population synthesis model employing a tidal prescription such as that developed here. 

A third explanation could be that some post-CE PN are not PN at all, but mimics \citep{Frew2010c}. Recently, \citet{Corradi2015} proposed that the Necklace PN, surrounding a post-CE central star binary, might not be a PN at all, in the sense that the nebula is not simply from the ejected AGB star envelope. This PN has a very low mass and collimated morphology. The proposal is that the PN around this binary has long gone and the current nebula is the product of an outburst that happened much later. If such a mechanism were common it could inflate the number of post-CE PN and possibly even explain the observed fraction of post-CE PN. However, in such case we would have to understand how the alternative mechanism operates and it is unclear whether it could operate often enough to justify the observed over-abundance of post-CE central stars of PN.  The suggestion of \citet{Corradi2015} could also be in line with the fact that about a quarter of all post-CE PN have an evolved companion, while predictions show that the incidence of evolved companions should be much lower than this \citep[e.g.,][]{Hillwig2004,DeMarco2015}.

\subsubsection{Summary of possible origin of the overabundance of observed post-CE central stars of PN}
\label{sssec:summary}
In summary here are the possible reasons why we observe a factor of several time more post-CE PN compared to what we expect using the binary fraction and period distribution of the progenitor population:
\begin{itemize}
\item The visibility time of post-CE PN is more than 6-8 times longer than the visibility time for PN from single and non-interacting binaries.
\item Not all single and non-interacting binary AGB stars make a visible PN. Common envelope binary interactions are a preferential channel to make PN. 
\item Some post-CE PN are not PN at all, but ejecta from a binary interaction that took place sometime after the PN has disappeared. This mechanism should also explain the over-abundance of double degenerate in the post-CE PN population.
\end{itemize}

\section*{Acknowledgments}

NM thanks Macquarie University Post-Graduate Research Fund and Prof. Dr. Ulrich Heber for financial help in travelling to EUROWD12 and ``Planetary Systems Beyond the Main Sequence" to present this work. NM thanks Prof. Miroslav Filipovic and Dr. Nick Tothill for their support.  We thank Alexander Mustill,  Mark Wardle and Jarrod Hurley for  helpful discussions. Falk Herwig and Jean-Claude Passy are acknowledged for his help with understanding MESA calculations of the upper AGB. Finally we thank an anonymous referee for helpful comments. OD acknowledges financial support from the Australian Research Council Future Fellowship scheme (FT120100452). EV hosted NM during his visit to Madrid by a grant provided by Marie Curie grant FP7-People-RG268111. EV acknowledges support  from the Spanish Ministerio de Econom\'{i}a y Competitividad under grant AYA2013-45347-P.  

\bibliography{../../../../../../Thesis/myrefs,/Users/orsola/Work/BibliographyFiles/bibliography}

\hyphenation{Post-Script Sprin-ger}
\begin{thebibliography}{}

\bibitem[\protect\citeauthoryear{{Adams} \& {Bloch}}{{Adams} \&
  {Bloch}}{2013}]{Adams2013}
{Adams} F.~C.,  {Bloch} A.~M.,  2013, \apjl, 777, L30

\bibitem[\protect\citeauthoryear{{Bertelli}, {Girardi}, {Marigo} \&
  {Nasi}}{{Bertelli} et~al.}{2008}]{Bertelli2008}
{Bertelli} G.,  {Girardi} L.,  {Marigo} P.,    {Nasi} E.,  2008, \aap, 484, 815

\bibitem[\protect\citeauthoryear{{Bertelli}, {Nasi}, {Girardi} \&
  {Marigo}}{{Bertelli} et~al.}{2009}]{Bertelli2009}
{Bertelli} G.,  {Nasi} E.,  {Girardi} L.,    {Marigo} P.,  2009, \aap, 508, 355

\bibitem[\protect\citeauthoryear{{Bloecker}}{{Bloecker}}{1995}]{Bloecker1995}
{Bloecker} T.,  1995, \aap, 299, 755

\bibitem[\protect\citeauthoryear{{Boffin} \& {Jorissen}}{{Boffin} \&
  {Jorissen}}{1988}]{Boffin1988}
{Boffin} H.~M.~J.,  {Jorissen} A.,  1988, \aap, 205, 155

\bibitem[\protect\citeauthoryear{{Bond}}{{Bond}}{2000}]{Bond2000}
{Bond} H.~E.,  2000, in ASP Conf. Ser. 199: Asymmetrical Planetary Nebulae II:
  From Origins to Microstructures {Binarity of Central Stars of Planetary
  Nebulae}.
p.~115

\bibitem[\protect\citeauthoryear{{Bondi} \& {Hoyle}}{{Bondi} \&
  {Hoyle}}{1944}]{Bondi1944}
{Bondi} H.,  {Hoyle} F.,  1944, \mnras, 104, 273

\bibitem[\protect\citeauthoryear{{Bressan}, {Marigo}, {Girardi}, {Salasnich},
  {Dal Cero}, {Rubele} \& {Nanni}}{{Bressan} et~al.}{2012}]{Bressan2012}
{Bressan} A.,  {Marigo} P.,  {Girardi} L.,  {Salasnich} B.,  {Dal Cero} C.,
  {Rubele} S.,    {Nanni} A.,  2012, \mnras, 427, 127

\bibitem[\protect\citeauthoryear{{Carlberg}, {Majewski} \& {Arras}}{{Carlberg}
  et~al.}{2009}]{Carlberg2009}
{Carlberg} J.~K.,  {Majewski} S.~R.,    {Arras} P.,  2009, \apj, 700, 832

\bibitem[\protect\citeauthoryear{{Chabrier}}{{Chabrier}}{2003}]{Chabrier2003big}
{Chabrier} G.,  2003, \pasp, 115, 763

\bibitem[\protect\citeauthoryear{{Claret}}{{Claret}}{2004}]{Claret2004}
{Claret} A.,  2004, \aap, 424, 919

\bibitem[\protect\citeauthoryear{{Corradi}, {Garc{\'{\i}}a-Rojas}, {Jones} \&
  {Rodr{\'{\i}}guez-Gil}}{{Corradi} et~al.}{2015}]{Corradi2015}
{Corradi} R.~L.~M.,  {Garc{\'{\i}}a-Rojas} J.,  {Jones} D.,
  {Rodr{\'{\i}}guez-Gil} P.,  2015, \apj, 803, 99

\bibitem[\protect\citeauthoryear{{Corradi} \& {Schwarz}}{{Corradi} \&
  {Schwarz}}{1995}]{Corradi1995}
{Corradi} R.~L.~M.,  {Schwarz} H.~E.,  1995, \aap, 293, 871

\bibitem[\protect\citeauthoryear{{De Marco}}{{De Marco}}{2009}]{DeMarco2009}
{De Marco} O.,  2009, \pasp, 121, 316

\bibitem[\protect\citeauthoryear{{De Marco}, {Hillwig} \& {Smith}}{{De Marco}
  et~al.}{2008}]{DeMarco2008}
{De Marco} O.,  {Hillwig} T.~C.,    {Smith} A.~J.,  2008, \aj, 136, 323

\bibitem[\protect\citeauthoryear{{De~Marco}, {Long}, {Jacoby}, {Hillwig},
  {Kronberger}, {Howell}, {Reindl} \& {Margheim}}{{De~Marco}
  et~al.}{2015}]{DeMarco2015}
{De~Marco} O.,  {Long} J.,  {Jacoby} G.~H.,  {Hillwig} T.,  {Kronberger} M.,
  {Howell} S.~B.,  {Reindl} N.,    {Margheim} S.,  2015, \mnras, 448, 3587

\bibitem[\protect\citeauthoryear{{De~Marco}, {Passy}, {Frew}, {Moe} \&
  {Jacoby}}{{De~Marco} et~al.}{2013}]{DeMarco2013}
{De~Marco} O.,  {Passy} J.-C.,  {Frew} D.~J.,  {Moe} M.,    {Jacoby} G.~H.,
  2013, \mnras, 428, 2118

\bibitem[\protect\citeauthoryear{{De~Marco}, {Passy}, {Moe}, {Herwig}, {Mac
  Low} \& {Paxton}}{{De~Marco} et~al.}{2011}]{DeMarco2011}
{De~Marco} O.,  {Passy} J.-C.,  {Moe} M.,  {Herwig} F.,  {Mac Low} M.-M.,
  {Paxton} B.,  2011, \mnras, 411, 2277

\bibitem[\protect\citeauthoryear{{Dijkstra} \& {Speck}}{{Dijkstra} \&
  {Speck}}{2006}]{Dijkstra2006}
{Dijkstra} C.,  {Speck} A.~K.,  2006, \apj, 651, 288

\bibitem[\protect\citeauthoryear{{Dorman}, {Rood} \& {O'Connell}}{{Dorman}
  et~al.}{1993}]{Dorman1993}
{Dorman} B.,  {Rood} R.~T.,    {O'Connell} R.~W.,  1993, \apj, 419, 596

\bibitem[\protect\citeauthoryear{{Douchin}, {De~Marco}, {Passy}, {Frew}, {Moe}
  \& {Jacoby}}{{Douchin} et~al.}{2015}]{Douchin2015}
{Douchin} D.,  {De~Marco} O.,  {Passy} J.-C.,  {Frew} D.~J.,  {Moe} M.,
  {Jacoby} G.~H.,  2015, \mnras, 428, 2118

\bibitem[\protect\citeauthoryear{{Duch{\^e}ne} \& {Kraus}}{{Duch{\^e}ne} \&
  {Kraus}}{2013}]{Duchene2013}
{Duch{\^e}ne} G.,  {Kraus} A.,  2013, \araa, 51, 269

\bibitem[\protect\citeauthoryear{{Duquennoy} \& {Mayor}}{{Duquennoy} \&
  {Mayor}}{1991}]{Duquennoy1991}
{Duquennoy} A.,  {Mayor} M.,  1991, \aap, 248, 485

\bibitem[\protect\citeauthoryear{{Eggleton}}{{Eggleton}}{1983}]{Eggleton1983}
{Eggleton} P.~P.,  1983, \apj, 268, 368

\bibitem[\protect\citeauthoryear{{Ekstr{\"o}m}, {Georgy}, {Eggenberger},
  {Meynet}, {Mowlavi}, {Wyttenbach}, {Granada}, {Decressin}, {Hirschi},
  {Frischknecht}, {Charbonnel} \& {Maeder}}{{Ekstr{\"o}m}
  et~al.}{2012}]{Ekstrom2012}
{Ekstr{\"o}m} S.,  {Georgy} C.,  {Eggenberger} P.,  {Meynet} G.,  {Mowlavi} N.,
   {Wyttenbach} A.,  {Granada} A.,  {Decressin} T.,  {Hirschi} R.,
  {Frischknecht} U.,  {Charbonnel} C.,    {Maeder} A.,  2012, \aap, 537, A146

\bibitem[\protect\citeauthoryear{{Farihi}, {Becklin} \& {Zuckerman}}{{Farihi}
  et~al.}{2005}]{Farihi2005}
{Farihi} J.,  {Becklin} E.~E.,    {Zuckerman} B.,  2005, \apjs, 161, 394

\bibitem[\protect\citeauthoryear{{Ferguson}, {Alexander}, {Allard}, {Barman},
  {Bodnarik}, {Hauschildt}, {Heffner-Wong} \& {Tamanai}}{{Ferguson}
  et~al.}{2005}]{Ferguson2005}
{Ferguson} J.~W.,  {Alexander} D.~R.,  {Allard} F.,  {Barman} T.,  {Bodnarik}
  J.~G.,  {Hauschildt} P.~H.,  {Heffner-Wong} A.,    {Tamanai} A.,  2005, \apj,
  623, 585

\bibitem[\protect\citeauthoryear{{Frew} \& {Parker}}{{Frew} \&
  {Parker}}{2010}]{Frew2010c}
{Frew} D.~J.,  {Parker} Q.~A.,  2010, PASA, 27, 129

\bibitem[\protect\citeauthoryear{{Garc{\'{\i}}a-Segura}, {Villaver}, {Langer},
  {Yoon} \& {Manchado}}{{Garc{\'{\i}}a-Segura} et~al.}{2014}]{GarciaSegura2014}
{Garc{\'{\i}}a-Segura} G.,  {Villaver} E.,  {Langer} N.,  {Yoon} S.-C.,
  {Manchado} A.,  2014, \apj, 783, 74

\bibitem[\protect\citeauthoryear{{Garc{\'{\i}}a-Segura}, {Villaver},
  {Manchado}, {Langer} \& {Yoon}}{{Garc{\'{\i}}a-Segura}
  et~al.}{2016}]{GarciaSegura2016}
{Garc{\'{\i}}a-Segura} G.,  {Villaver} E.,  {Manchado} A.,  {Langer} N.,
  {Yoon} S.-C.,  2016, \apj, 823, 142

\bibitem[\protect\citeauthoryear{{Girardi}, {Bressan}, {Bertelli} \&
  {Chiosi}}{{Girardi} et~al.}{2000}]{Girardi2000}
{Girardi} L.,  {Bressan} A.,  {Bertelli} G.,    {Chiosi} C.,  2000, \aaps, 141,
  371

\bibitem[\protect\citeauthoryear{{Goldreich} \& {Keeley}}{{Goldreich} \&
  {Keeley}}{1977}]{Goldreich1977}
{Goldreich} P.,  {Keeley} D.~A.,  1977, \apj, 211, 934

\bibitem[\protect\citeauthoryear{{Han}, {Podsiadlowski} \& {Eggleton}}{{Han}
  et~al.}{1995}]{Han1995}
{Han} Z.,  {Podsiadlowski} P.,    {Eggleton} P.~P.,  1995, \mnras, 272, 800

\bibitem[\protect\citeauthoryear{{Herwig}}{{Herwig}}{2000}]{Herwig2000}
{Herwig} F.,  2000, \aap, 360, 952

\bibitem[\protect\citeauthoryear{{Herwig} \& {Austin}}{{Herwig} \&
  {Austin}}{2004}]{Herwig2004}
{Herwig} F.,  {Austin} S.~M.,  2004, \apjl, 613, L73

\bibitem[\protect\citeauthoryear{{Hillwig}}{{Hillwig}}{2004}]{Hillwig2004}
{Hillwig} T.~C.,  2004, in {Meixner} M.,  {Kastner} J.~H.,  {Balick} B.,
  {Soker} N.,  eds, Asymmetrical Planetary Nebulae III: Winds, Structure and
  the Thunderbird Vol.~313 of Astronomical Society of the Pacific Conference
  Series, {Two New Close Binary Central Stars of Planetary Nebulae from a
  Critically Selected Southern Hemisphere Sample}.
p.~529

\bibitem[\protect\citeauthoryear{{Hillwig}, {Bond}, {Af{\c s}ar} \&
  {De~Marco}}{{Hillwig} et~al.}{2010}]{Hillwig2010}
{Hillwig} T.~C.,  {Bond} H.~E.,  {Af{\c s}ar} M.,    {De~Marco} O.,  2010, \aj,
  140, 319

\bibitem[\protect\citeauthoryear{{Hillwig}, {Frew}, {Louie}, {De Marco},
  {Bond}, {Jones} \& {Schaub}}{{Hillwig} et~al.}{2015}]{Hillwig2015}
{Hillwig} T.~C.,  {Frew} D.~J.,  {Louie} M.,  {De Marco} O.,  {Bond} H.~E.,
  {Jones} D.,    {Schaub} S.~C.,  2015, \aj, 150, 30

\bibitem[\protect\citeauthoryear{{Huarte-Espinosa}, {Carroll-Nellenback},
  {Nordhaus}, {Frank} \& {Blackman}}{{Huarte-Espinosa}
  et~al.}{2013}]{HuarteEspinosa2013}
{Huarte-Espinosa} M.,  {Carroll-Nellenback} J.,  {Nordhaus} J.,  {Frank} A.,
  {Blackman} E.~G.,  2013, \mnras, 433, 295

\bibitem[\protect\citeauthoryear{{Hurley}, {Pols} \& {Tout}}{{Hurley}
  et~al.}{2000}]{Hurley2000}
{Hurley} J.~R.,  {Pols} O.~R.,    {Tout} C.~A.,  2000, \mnras, 315, 543

\bibitem[\protect\citeauthoryear{{Hurley}, {Tout} \& {Pols}}{{Hurley}
  et~al.}{2002}]{Hurley2002}
{Hurley} J.~R.,  {Tout} C.~A.,    {Pols} O.~R.,  2002, \mnras, 329, 897

\bibitem[\protect\citeauthoryear{{Iben} Jr. \& {Tutukov}}{{Iben} \&
  {Tutukov}}{1985}]{Iben1985}
{Iben} Jr. I.,  {Tutukov} A.~V.,  1985, \apjs, 58, 661

\bibitem[\protect\citeauthoryear{{Ireland}, {Scholz} \& {Wood}}{{Ireland}
  et~al.}{2011}]{Ireland2011}
{Ireland} M.~J.,  {Scholz} M.,    {Wood} P.~R.,  2011, \mnras, 418, 114

\bibitem[\protect\citeauthoryear{{Ivanova}, {Justham}, {Chen}, {De Marco},
  {Fryer}, {Gaburov}, {Ge}, {Glebbeek}, {Han}, {Li}, {Lu}, {Marsh},
  {Podsiadlowski}, {Potter}, {Soker}, {Taam}, {Tauris}, {van den Heuvel} \&
  {Webbink}}{{Ivanova} et~al.}{2013}]{Ivanova2013}
{Ivanova} N.,  {Justham} S.,  {Chen} X.,  {De Marco} O.,  {Fryer} C.~L.,
  {Gaburov} E.,  {Ge} H.,  {Glebbeek} E.,  {Han} Z.,  {Li} X.-D.,  {Lu} G.,
  {Marsh} T.,  {Podsiadlowski} P.,  {Potter} A.,  {Soker} N.,  {Taam} R.,
  {Tauris} T.~M.,  {van den Heuvel} E.~P.~J.,    {Webbink} R.~F.,  2013, \aapr,
  21, 59

\bibitem[\protect\citeauthoryear{{Izzard}, {Glebbeek}, {Stancliffe} \&
  {Pols}}{{Izzard} et~al.}{2009}]{Izzard2009}
{Izzard} R.~G.,  {Glebbeek} E.,  {Stancliffe} R.~J.,    {Pols} O.~R.,  2009,
  \aap, 508, 1359

\bibitem[\protect\citeauthoryear{{Jacoby}, {Kronberger}, {Patchick}, {Teutsch},
  {Saloranta}, {Howell}, {Crisp}, {Riddle}, {Acker}, {Frew} \&
  {Parker}}{{Jacoby} et~al.}{2010}]{Jacoby2010}
{Jacoby} G.~H.,  {Kronberger} M.,  {Patchick} D.,  {Teutsch} P.,  {Saloranta}
  J.,  {Howell} M.,  {Crisp} R.,  {Riddle} D.,  {Acker} A.,  {Frew} D.~J.,
  {Parker} Q.~A.,  2010, \pasa, 27, 156

\bibitem[\protect\citeauthoryear{{Jacoby}, {Morse}, {Fullton}, {Kwitter} \&
  {Henry}}{{Jacoby} et~al.}{1997}]{Jacoby1997}
{Jacoby} G.~H.,  {Morse} J.~A.,  {Fullton} L.~K.,  {Kwitter} K.~B.,    {Henry}
  R.~B.~C.,  1997, \aj, 114, 2611

\bibitem[\protect\citeauthoryear{{Jones}, {Boffin}, {Miszalski}, {Wesson},
  {Corradi} \& {Tyndall}}{{Jones} et~al.}{2014}]{Jones2014}
{Jones} D.,  {Boffin} H.~M.~J.,  {Miszalski} B.,  {Wesson} R.,  {Corradi}
  R.~L.~M.,    {Tyndall} A.~A.,  2014, \aap, 562, A89

\bibitem[\protect\citeauthoryear{{Karakas}, {Lattanzio} \& {Pols}}{{Karakas}
  et~al.}{2002}]{Karakas2002}
{Karakas} A.~I.,  {Lattanzio} J.~C.,    {Pols} O.~R.,  2002, \pasa, 19, 515

\bibitem[\protect\citeauthoryear{{Kingsburgh} \& {Barlow}}{{Kingsburgh} \&
  {Barlow}}{1994}]{Kingsburgh1994}
{Kingsburgh} R.~L.,  {Barlow} M.~J.,  1994, \mnras, 271, 257

\bibitem[\protect\citeauthoryear{{Liebert}, {Bergeron} \& {Holberg}}{{Liebert}
  et~al.}{2005}]{Liebert2005}
{Liebert} J.,  {Bergeron} P.,    {Holberg} J.~B.,  2005, \apjs, 156, 47

\bibitem[\protect\citeauthoryear{{Marigo}, {Bressan}, {Nanni}, {Girardi} \&
  {Pumo}}{{Marigo} et~al.}{2013}]{Marigo2013}
{Marigo} P.,  {Bressan} A.,  {Nanni} A.,  {Girardi} L.,    {Pumo} M.~L.,  2013,
  \mnras, 434, 488

\bibitem[\protect\citeauthoryear{{McDonald} \& {Zijlstra}}{{McDonald} \&
  {Zijlstra}}{2015}]{McDonald2015}
{McDonald} I.,  {Zijlstra} A.~A.,  2015, \mnras, 448, 502

\bibitem[\protect\citeauthoryear{{Miller Bertolami}}{{Miller
  Bertolami}}{2016}]{MillerBertolami2016}
{Miller Bertolami} M.~M.,  2016, \aap, 588, A25

\bibitem[\protect\citeauthoryear{{Miszalski}, {Acker}, {Moffat}, {Parker} \&
  {Udalski}}{{Miszalski} et~al.}{2009}]{Miszalski2009}
{Miszalski} B.,  {Acker} A.,  {Moffat} A.~F.~J.,  {Parker} Q.~A.,    {Udalski}
  A.,  2009, \aap, 496, 813

\bibitem[\protect\citeauthoryear{{Miszalski}, {Acker}, {Parker} \&
  {Moffat}}{{Miszalski} et~al.}{2009}]{Miszalski2009b}
{Miszalski} B.,  {Acker} A.,  {Parker} Q.~A.,    {Moffat} A.~F.~J.,  2009,
  \aap, 505, 249

\bibitem[\protect\citeauthoryear{{Mitchell}, {Pollacco}, {O'Brien}, {Bryce},
  {L{\'o}pez}, {Meaburn} \& {Vaytet}}{{Mitchell} et~al.}{2007}]{Mitchell2007}
{Mitchell} D.~L.,  {Pollacco} D.,  {O'Brien} T.~J.,  {Bryce} M.,  {L{\'o}pez}
  J.~A.,  {Meaburn} J.,    {Vaytet} N.~M.~H.,  2007, \mnras, 374, 1404

\bibitem[\protect\citeauthoryear{{Moe} \& {De Marco}}{{Moe} \& {De
  Marco}}{2006}]{MoeDeMarco2006}
{Moe} M.,  {De Marco} O.,  2006, \apj, 650, 916

\bibitem[\protect\citeauthoryear{{Moe} \& {De~Marco}}{{Moe} \&
  {De~Marco}}{2006}]{Moe2006}
{Moe} M.,  {De~Marco} O.,  2006, \apj, 650, 916

\bibitem[\protect\citeauthoryear{{Moe} \& {De Marco}}{{Moe} \& {De
  Marco}}{2012}]{Moe2012}
{Moe} M.,  {De Marco} O.,  2012, in IAU Symposium Vol.~283 of IAU Symposium,
  {Population synthesis of planetary nebulae from binaries}.
pp 111--114

\bibitem[\protect\citeauthoryear{{Mustill} \& {Villaver}}{{Mustill} \&
  {Villaver}}{2012}]{Mustill2012}
{Mustill} A.~J.,  {Villaver} E.,  2012, \apj, 761, 121

\bibitem[\protect\citeauthoryear{{Nie}, {Wood} \& {Nicholls}}{{Nie}
  et~al.}{2012}]{Nie2012}
{Nie} J.~D.,  {Wood} P.~R.,    {Nicholls} C.~P.,  2012, \mnras, 423, 2764

\bibitem[\protect\citeauthoryear{{Nordhaus} \& {Blackman}}{{Nordhaus} \&
  {Blackman}}{2006}]{Nordhaus2006}
{Nordhaus} J.,  {Blackman} E.~G.,  2006, \mnras, 370, 2004

\bibitem[\protect\citeauthoryear{{Nordhaus}, {Blackman} \& {Frank}}{{Nordhaus}
  et~al.}{2007}]{Nordhaus2007}
{Nordhaus} J.,  {Blackman} E.~G.,    {Frank} A.,  2007, \mnras, 376, 599

\bibitem[\protect\citeauthoryear{{Nordhaus} \& {Spiegel}}{{Nordhaus} \&
  {Spiegel}}{2013}]{Nordhaus2013}
{Nordhaus} J.,  {Spiegel} D.~S.,  2013, \mnras

\bibitem[\protect\citeauthoryear{{Paczynski}}{{Paczynski}}{1976}]{Paczynski1976}
{Paczynski} B.,  1976, in {P.~Eggleton, S.~Mitton, \& J.~Whelan} ed., Structure
  and Evolution of Close Binary Systems Vol.~73 of IAU Symposium, {Common
  Envelope Binaries}.
pp 75--+

\bibitem[\protect\citeauthoryear{{Parker}, {Acker}, {Frew}, {Hartley},
  {Peyaud}, {Ochsenbein}, {Phillipps}, {Russeil}, {Beaulieu}, {Cohen},
  {K{\"o}ppen}, {Miszalski}, {Morgan}, {Morris}, {Pierce} \&
  {Vaughan}}{{Parker} et~al.}{2006}]{Parker2006}
{Parker} Q.~A.,  {Acker} A.,  {Frew} D.~J.,  {Hartley} M.,  {Peyaud} A.~E.~J.,
  {Ochsenbein} F.,  {Phillipps} S.,  {Russeil} D.,  {Beaulieu} S.~F.,  {Cohen}
  M.,  {K{\"o}ppen} J.,  {Miszalski} B.,  {Morgan} D.~H.,  {Morris} R.~A.~H.,
  {Pierce} M.~J.,    {Vaughan} A.~E.,  2006, \mnras, 373, 79

\bibitem[\protect\citeauthoryear{{Passy}, {Herwig} \& {Paxton}}{{Passy}
  et~al.}{2012a}]{Passy2012}
{Passy} J.-C.,  {Herwig} F.,    {Paxton} B.,  2012a, \apj, 760, 90

\bibitem[\protect\citeauthoryear{{Passy}, {Herwig} \& {Paxton}}{{Passy}
  et~al.}{2012b}]{Passy2012b}
{Passy} J.-C.,  {Herwig} F.,    {Paxton} B.,  2012b, \apj, 760, 90

\bibitem[\protect\citeauthoryear{{Paxton}, {Bildsten}, {Dotter}, {Herwig},
  {Lesaffre} \& {Timmes}}{{Paxton} et~al.}{2011}]{Paxton2011}
{Paxton} B.,  {Bildsten} L.,  {Dotter} A.,  {Herwig} F.,  {Lesaffre} P.,
  {Timmes} F.,  2011, \apjs, 192, 3

\bibitem[\protect\citeauthoryear{{Paxton}, {Cantiello}, {Arras}, {Bildsten},
  {Brown}, {Dotter}, {Mankovich}, {Montgomery}, {Stello}, {Timmes} \&
  {Townsend}}{{Paxton} et~al.}{2013}]{Paxton2013}
{Paxton} B.,  {Cantiello} M.,  {Arras} P.,  {Bildsten} L.,  {Brown} E.~F.,
  {Dotter} A.,  {Mankovich} C.,  {Montgomery} M.~H.,  {Stello} D.,  {Timmes}
  F.~X.,    {Townsend} R.,  2013, ArXiv e-prints

\bibitem[\protect\citeauthoryear{{Penev}, {Zhang} \& {Jackson}}{{Penev}
  et~al.}{2014}]{Penev2014}
{Penev} K.,  {Zhang} M.,    {Jackson} B.,  2014, \pasp, 126, 553

\bibitem[\protect\citeauthoryear{{Pols}, {Schr{\"o}der}, {Hurley}, {Tout} \&
  {Eggleton}}{{Pols} et~al.}{1998}]{Pols1998}
{Pols} O.~R.,  {Schr{\"o}der} K.-P.,  {Hurley} J.~R.,  {Tout} C.~A.,
  {Eggleton} P.~P.,  1998, \mnras, 298, 525

\bibitem[\protect\citeauthoryear{{Pols}, {Tout}, {Lattanzio} \&
  {Karakas}}{{Pols} et~al.}{2001}]{Pols2001}
{Pols} O.~R.,  {Tout} C.~A.,  {Lattanzio} J.~C.,    {Karakas} A.~I.,  2001, in
  {Podsiadlowski} P.,  {Rappaport} S.,  {King} A.~R.,  {D'Antona} F.,
  {Burderi} L.,  eds, Evolution of Binary and Multiple Star Systems Vol.~229 of
  Astronomical Society of the Pacific Conference Series, {Thermal Pulses and
  Dredge-up in AGB Stars}.
p.~31

\bibitem[\protect\citeauthoryear{{Privitera}, {Meynet}, {Eggenberger},
  {Vidotto}, {Villaver} \& {Bianda}}{{Privitera} et~al.}{2016}]{Privitera2016}
{Privitera} G.,  {Meynet} G.,  {Eggenberger} P.,  {Vidotto} A.~A.,  {Villaver}
  E.,    {Bianda} M.,  2016, ArXiv e-prints

\bibitem[\protect\citeauthoryear{{Raghavan}, {McAlister}, {Henry}, {Latham},
  {Marcy}, {Mason}, {Gies}, {White} \& {ten Brummelaar}}{{Raghavan}
  et~al.}{2010}]{Raghavan2010}
{Raghavan} D.,  {McAlister} H.~A.,  {Henry} T.~J.,  {Latham} D.~W.,  {Marcy}
  G.~W.,  {Mason} B.~D.,  {Gies} D.~R.,  {White} R.~J.,    {ten Brummelaar}
  T.~A.,  2010, \apjs, 190, 1

\bibitem[\protect\citeauthoryear{{Rasio}, {Tout}, {Lubow} \& {Livio}}{{Rasio}
  et~al.}{1996}]{Rasio1996}
{Rasio} F.~A.,  {Tout} C.~A.,  {Lubow} S.~H.,    {Livio} M.,  1996, \apj, 470,
  1187

\bibitem[\protect\citeauthoryear{{Reimers}}{{Reimers}}{1975}]{Reimers1975}
{Reimers} D.,  1975, Memoires of the Societe Royale des Sciences de Liege, 8,
  369

\bibitem[\protect\citeauthoryear{{Siess}}{{Siess}}{2006}]{Siess2006}
{Siess} L.,  2006, \aap, 448, 717

\bibitem[\protect\citeauthoryear{{Soker}}{{Soker}}{1994}]{Soker1994c}
{Soker} N.,  1994, \pasp, 106, 59

\bibitem[\protect\citeauthoryear{{Soker}}{{Soker}}{1996}]{Soker1996}
{Soker} N.,  1996, \apjl, 460, L53+

\bibitem[\protect\citeauthoryear{{Soker}}{{Soker}}{1997}]{Soker1997}
{Soker} N.,  1997, \apj, 112, 487

\bibitem[\protect\citeauthoryear{{Soker}}{{Soker}}{1998}]{Soker1998}
{Soker} N.,  1998, \apj, 496, 833

\bibitem[\protect\citeauthoryear{{Soker}}{{Soker}}{2006}]{Soker2006}
{Soker} N.,  2006, \pasp, 118, 260

\bibitem[\protect\citeauthoryear{{Soker} \& {Subag}}{{Soker} \&
  {Subag}}{2005}]{Soker2005}
{Soker} N.,  {Subag} E.,  2005, \aj, 130, 2717

\bibitem[\protect\citeauthoryear{{Stancliffe}, {Fossati}, {Passy} \&
  {Schneider}}{{Stancliffe} et~al.}{2016}]{Stancliffe2016}
{Stancliffe} R.~J.,  {Fossati} L.,  {Passy} J.-C.,    {Schneider} F.~R.~N.,
  2016, \aap, 586, A119

\bibitem[\protect\citeauthoryear{{Stasinska}, {Gorny} \& {Tylenda}}{{Stasinska}
  et~al.}{1997}]{Stasinska1997}
{Stasinska} G.,  {Gorny} S.~K.,    {Tylenda} R.,  1997, \aap, 327, 736

\bibitem[\protect\citeauthoryear{{Sterne}}{{Sterne}}{1939}]{Sterne1939}
{Sterne} T.~E.,  1939, \mnras, 99, 451

\bibitem[\protect\citeauthoryear{{Torres}, {Andersen} \&
  {Gim{\'e}nez}}{{Torres} et~al.}{2010}]{Torres2010}
{Torres} G.,  {Andersen} J.,    {Gim{\'e}nez} A.,  2010, \aapr, 18, 67

\bibitem[\protect\citeauthoryear{{Tout} \& {Eggleton}}{{Tout} \&
  {Eggleton}}{1988}]{Tout1988}
{Tout} C.~A.,  {Eggleton} P.~P.,  1988, \mnras, 231, 823

\bibitem[\protect\citeauthoryear{{Trampedach} \& {Stein}}{{Trampedach} \&
  {Stein}}{2011}]{Trampedach2011}
{Trampedach} R.,  {Stein} R.~F.,  2011, \apj, 731, 78

\bibitem[\protect\citeauthoryear{{Vassiliadis} \& {Wood}}{{Vassiliadis} \&
  {Wood}}{1994}]{Vassiliadis1994}
{Vassiliadis} E.,  {Wood} P.~R.,  1994, \apjs, 92, 125

\bibitem[\protect\citeauthoryear{{Ventura}, {Criscienzo}, {Schneider},
  {Carini}, {Valiante}, {D'Antona}, {Gallerani}, {Maiolino} \&
  {Tornamb{\'e}}}{{Ventura} et~al.}{2012}]{Ventura2012}
{Ventura} P.,  {Criscienzo} M.~D.,  {Schneider} R.,  {Carini} R.,  {Valiante}
  R.,  {D'Antona} F.,  {Gallerani} S.,  {Maiolino} R.,    {Tornamb{\'e}} A.,
  2012, \mnras, 424, 2345

\bibitem[\protect\citeauthoryear{{Ventura}, {Di Criscienzo}, {Carini} \&
  {D'Antona}}{{Ventura} et~al.}{2013}]{Ventura2013}
{Ventura} P.,  {Di Criscienzo} M.,  {Carini} R.,    {D'Antona} F.,  2013,
  \mnras, 431, 3642

\bibitem[\protect\citeauthoryear{{Verbunt} \& {Phinney}}{{Verbunt} \&
  {Phinney}}{1995}]{Verbunt1995}
{Verbunt} F.,  {Phinney} E.~S.,  1995, \aap, 296, 709

\bibitem[\protect\citeauthoryear{{Villaver} \& {Livio}}{{Villaver} \&
  {Livio}}{2009}]{Villaver2009}
{Villaver} E.,  {Livio} M.,  2009, \apjl, 705, L81

\bibitem[\protect\citeauthoryear{{Villaver}, {Livio}, {Mustill} \&
  {Siess}}{{Villaver} et~al.}{2014}]{Villaver2014}
{Villaver} E.,  {Livio} M.,  {Mustill} A.~J.,    {Siess} L.,  2014, \apj, 794,
  3

\bibitem[\protect\citeauthoryear{{Zahn}}{{Zahn}}{1989}]{Zahn1989}
{Zahn} J.,  1989, \aap, 220, 112

\bibitem[\protect\citeauthoryear{{Zahn}}{{Zahn}}{1977}]{Zahn1977}
{Zahn} J.~P.,  1977, \aap, 57, 383

\end{thebibliography}

\bibliographystyle{mn2e}
\label{lastpage}

\end{document}